\documentclass[superscriptaddress, twocolumn, prb, mediabb, nobibnotes]{revtex4}

\usepackage{amssymb,amsmath,array,verbatim}
\usepackage[dvipdfmx]{graphicx}

\newcommand{\beq}{\begin{eqnarray}}
\newcommand{\eeq}{\end{eqnarray}}

\newcommand{\bpm}{\begin{pmatrix}}
\newcommand{\epm}{\end{pmatrix}}

\newcommand{\ba}{\left(\begin{array}}
\newcommand{\ea}{\end{array} \right)}

\begin{document}

\title{New class of flat-band models on tetragonal and hexagonal 
lattices:\\
Gapped versus crossing flat bands}

\author{Tatsuhiro Misumi}
\email{misumi@phys.akita-u.ac.jp}
\address{Department of Mathematical Science, Akita University, Akita 010-8502, Japan}
\address{Research and Education Center for Natural Sciences, Keio University, Kanagawa 223-8521, Japan}
\address{Interdisciplinary Theoretical and Mathematical Sciences Program, RIKEN, Saitama 351-0198, Japan}

\author{Hideo Aoki}
\email[Now also at Department of Physics, ETH Z\"{u}rich, 8093 Z\"{u}rich, Switzerland; ]{aoki@phys.s.u-tokyo.ac.jp}
\address{Department of Physics, The University of Tokyo, Hongo, Tokyo 113-0033, Japan }
\address{Electronics and Photonics Research Institute, Advanced Industrial Science and Technology (AIST), Tsukuba, Ibaraki 305-8568, Japan}

\begin{abstract}
We propose a new class of tight-binding models 
where a flat band exists either gapped from or crossing right through a dispersive band 
on two-band (i.e., two sites/unit cell) tetragonal and honeycomb lattices. 
By imposing a condition on the hopping parameters for 
generic models with up to third-neighbor hoppings, 
we first obtain models having a rigorously flat band isolated 
from a dispersive band with a gap, which accommodate both rank-reducing 
and non-rank-reducing of the Hamiltonian.  
The models include Tasaki's flat-band models, 
but the present model generally 
has a nonzero flat-band energy whose gap from the dispersive band is controllable as well.
We then modify the models by appropriately 
changing the second- or third-neighbor hoppings, leading to a new class of 
two-dimensional lattices where a (slightly warped) flat band 
pierces a dispersive one. As with the known flat-band models, 
the connectivity condition is satisfied in the present models, 
so that we have unusual, unorthogonalizable Wannier orbitals. 
We have also shown that the present flat-band model can be extended to 
three (or higher) dimensions.
Implications on possible high-$T_{C}$ superconductivity are discussed 
when a repulsive electron-electron interaction is 
introduced, where the flat band is envisaged to be 
utilized as intermediate states in pair scattering processes.  

\end{abstract}

\maketitle

\section{Introduction}

Since the late 1980's, it has been recognized that electronic 
energy bands can become dispersionless due to quantum interference 
in certain classes of multi-band lattice systems, 
which is totally different from a trivial atomic limit with vanishing hopping.  
Such systems, called ``flat-band" systems, were 
kicked off by Lieb\cite{Lieb:1989}, 
which was subsequently extended by Mielke and by Tasaki\cite{Mielke:1991,Tasaki:1992,Tasaki:1998}.
In these models the interests were magnetism, 
for which ferromagnetism is rigorously proved for arbitrary onsite (Hubbard) repulsion
when the flat band is half-filled. The mechanism for 
the magnetism is that we have an unusual situation where 
Wannier orbitals cannot be orthogonalized, so that 
Pauli exclusion minimizes the repulsive interaction when spins 
are aligned\cite{Tasaki:1992}. 

In quite a different avenue, people have started to 
explore whether flat-band models could accommodate unconventional superconductivity\cite{Kuroki:2005kha,Takayoshi:2013tkw,Tovmasyan:2013tnh,Kobayashi:2016koy}. Specifically, when the electron-electron interaction is repulsive, 
virtual pair scatterings between the dispersive 
and flat bands in the spin-fluctuation meditation pairing 
are recognized to lead to relatively high-$T_{C}$ superconductivity
when the flat and dispersive bands are 
intersecting with each other\cite{Kuroki:2005kha,Takayoshi:2013tkw,Tovmasyan:2013tnh,Kobayashi:2016koy}.  
The flat-band system has also attracted attention in the context of topological properties, 
since a nontrivial Chern number can be defined when a flat band is separated 
from other ones \cite{Tang:2011tmw,Sun:2011sgk,Neupert:2011nsc,Li:2013lzl}.  
Superconducting and topological properties may even be related with 
each other: Recently T\"{o}rma and coworkers have examined the superfluid weight 
for the flat-band superconductivity in terms of the Chern number for attractive 
interactions\cite{Peotta:2015pto, Tovmasyan:2016:tpt}.  For repulsive interactions, 
Kobayashi {\it et al}. show that the 
superconductivity in a flat-band model sits right next to a 
topological-insulator phase\cite{Kobayashi:2016koy}. 
Orbital susceptibility in the flat-band models 
is also discussed in Refs.\cite{AokiAndo96,Raoux:2014rmf,Piechon:2016prf}.
These studies attest that flat-band systems can be of great importance in 
condensed-matter physics in terms of magnetism, superconductivity and topological properties.

So far, three classes of flat-band models have been proposed: Lieb, Mielke and Tasaki models\cite{Lieb:1989,Mielke:1991,Tasaki:1992}.  
In Lieb's bipartite models \cite{Lieb:1989}, the flat band has exactly 
zero energy as imposed by the electron-hole symmetry.  
The emergence of flat bands in the model is due to the fact that the rank of the tight-binding Hamiltonian matrix for the AB bipartite model is reduced when the total number of A-sublattice sites differs from that for B sites.  
By contrast Mielke's model \cite{Mielke:1991} is constructed based on a ``line-graph" method in the graph theory, 
which can also give rise to dispersionless bands.  
A typical Mielke model is kagome lattice, which is a line graph of honeycomb lattice. 
Tasaki's model \cite{Tasaki:1992} is built in a systematic manner called a ``cell construction",
where cells, each composed of a single internal site and $n$ external sites, 
are connected with $m$ cells sharing an external site [called $(n,m)$-Tasaki model].  
Unlike Lieb's and Mielke's models, Tasaki's model induces a flat band that is 
separated from a dispersive one with a nonzero energy gap.  
There, the flat band arises via reduction of the rank of the Hamiltonian, 
which imposes the energy of the flat band to be zero.

Now, if the expectation that a flat band intersecting a dispersive one 
can favor superconductivity is correct, it is highly desirable to 
{\it systematically} construct a class of flat-band models where 
the intersection is controllable, most simply in two-band models in two (or higher) dimensions.  
In this work, we precisely propose such a new class of flat-band 
models. 
We start with constructing a case of isolated flat band with tunable energy 
for tight-binding models, on two-dimensional 
body-centered tetragonal lattice and honeycomb lattice with up to the third-neighbor hoppings. 
We then impose an  ``auxiliary condition" to 
obtain exactly flat band models composed of both non-rank-reducing and rank-reducing cases of the Hamiltonian, where Wannier orbitals are still unorthogonalizable.  An advantage of thus obtained class of models 
is that the flat band has a nonzero energy 
(as in Mielke's model) with a gap between the flat and dispersive bands 
(as in Tasaki's model), where the gap is controllable.    
The model is then 
used to construct a case of crossing flat and dispersive bands 
by tuning the second- or third-neighbor hoppings,  
which realizes the flat band piercing right through the dispersive band, although 
the flat band becomes slightly warped.  
We shall finally show that the present flat-band models can be 
readily extended to higher dimensions.

This paper is organized as follows:
In Sec.~\ref{sec:tetragonal}, we construct a flat-band system on the tetragonal lattice.
We then consider in Sec.~\ref{sec:tetragonald} a one-parameter deformation of the tetragonal flat-band model
to realize the flat band piercing the dispersive band.  
Section~\ref{sec:honey} constructs a flat-band model on the honeycomb lattice.  
In Sec.~\ref{sec:bcc} and Appendix~\ref{sec:bccD}, respectively, 
we construct a flat-band model on 
three-dimensional (3D) and general-dimensional body-centered-cubic lattices.  
Section~\ref{sec:SD} is devoted to summary and discussions.


\section{Flat-band model on tetragonal lattices: gapped case}
\label{sec:tetragonal}

\subsection{Model}

\begin{figure}[htbp]
\begin{center}
\includegraphics[width=0.45\textwidth]{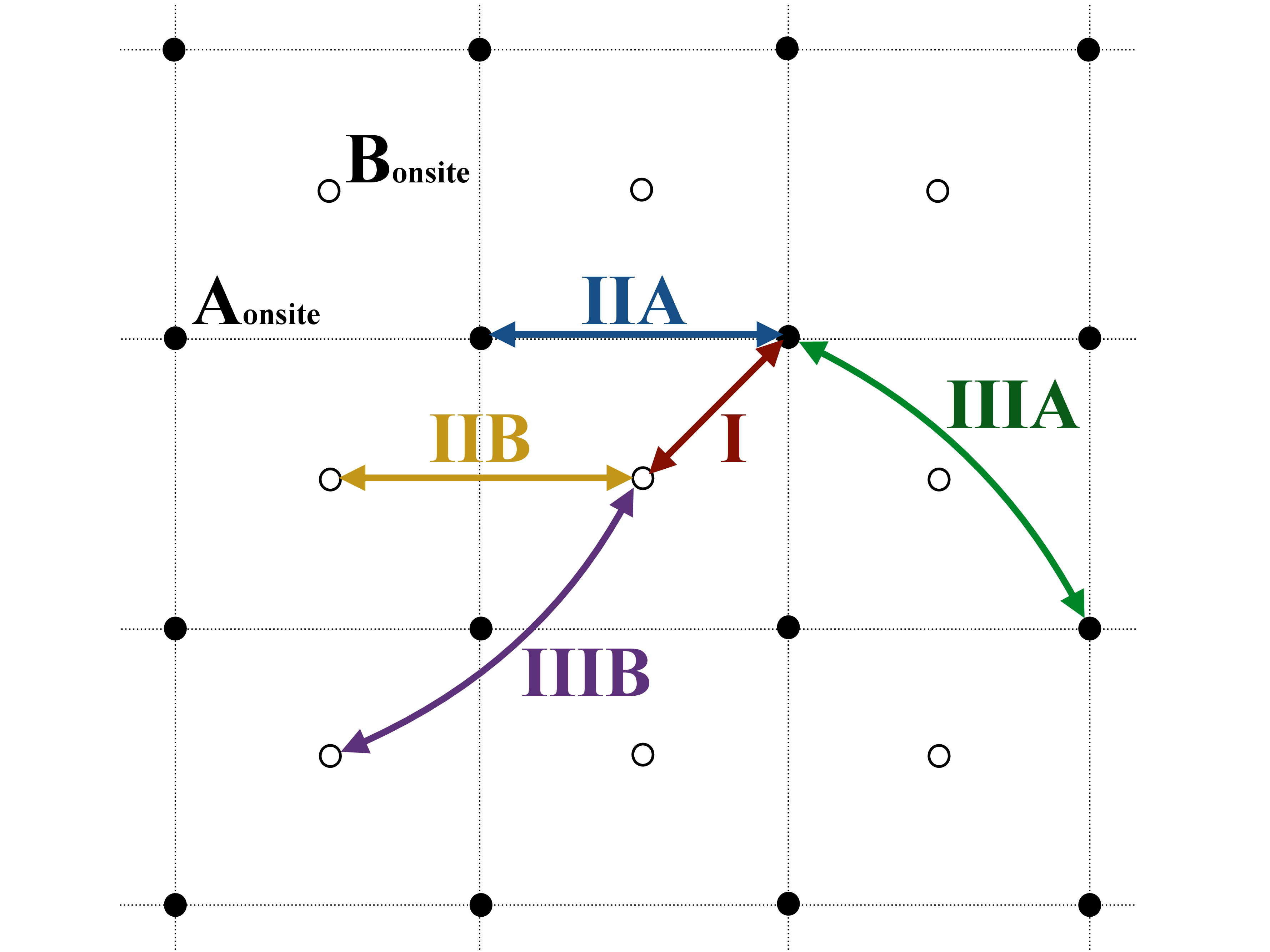}
\end{center}
\caption{The present lattice model, where 
the hoppings (arrows) are given by Eq.(\ref{eq:hoptetragonal}) in the text.
Solid and open circles represent A and B sublattice sites, respectively.}
\label{fig:sites_tetragonal}
\end{figure}

We begin with a single-particle tight-binding model, 
\[
H\,=\, \sum_{i,j}t_{i,j}c_{i}^{\dag}c_{j}, 
\]
where $t_{i,j} (i\neq j)$ is the transfer energy, $t_{i,i}$ the 
onsite energy, and $c_{i}^{\dag}$ creates an electron at site $i$ 
with suppressed spin indices.  We consider a lattice, here 
exemplified by a tetragonal one, that can be 
divided into two (A and B) sublattices with two sites per 
unit cell, which implies we are considering a two-band model.  
The model, depicted in Fig.~\ref{fig:sites_tetragonal}, 
comprises nearest-neighbor($t$), second-neighbor 
($t^{(2)}$) and 
third-neighbor ($t^{(3)}$) hoppings along with 
onsite energy ($t^{(0)}$) as
\begin{align}
&{\rm A_{onsite}}\,\,:\,\, t_{\rm AA}^{(0)} = a,\quad
&{\rm B_{onsite}}\,\,:\,\, t_{\rm BB}^{(0)} = b, \nonumber\\
&{\rm I}\,\,:\,\,\,\quad t_{\rm AB} = c, \nonumber\\
&{\rm IIA}\,\,:\,\,t_{\rm AA}^{(2)} = d,\quad
&{\rm IIIA}\,\,:\,\, t_{\rm AA}^{(3)} = {d\over2}, \nonumber\\
&{\rm IIB}\,\,:\,\, t_{\rm BB}^{(2)} = e,\quad
&{\rm IIIB}\,\, : \,\, t_{\rm BB}^{(3)} = {e\over2}, \nonumber\\
\label{eq:hoptetragonal}
\end{align}
where subscripts denote sublattices and $a,b,c,d,e \in {\mathbb R}$.
Here, $c$ is the nearest hopping between AB, 
while $d(e)$ characterizes AA (BB) hoppings, where 
the next-nearest hopping on each (A or B) sublattice is set to be 
half the nearest one on the sublattice.

In $k$-space the Hamiltonian is
\begin{align}
{\mathcal H}=
\begin{pmatrix}
{\mathcal H}_{\rm AA}  & {\mathcal H}_{\rm AB} \\
{\mathcal H}_{\rm BA} & {\mathcal H}_{\rm BB}
\end{pmatrix}\,,
\label{eq:1Htetragonal}
\end{align}
with
\begin{align}
&{\mathcal H}_{\rm AA}  = a\,+\,  2d\,F(k_{x},k_{y})\,,
\nonumber\\
&{\mathcal H}_{\rm AB} = {\mathcal H}_{\rm BA} = 4c\cos{k_{x}\over{2}}\cos {k_{y}\over{2}} \,,
\nonumber\\
&{\mathcal H}_{\rm BB} = b\,+\, 2e\,F(k_{x},k_{y})\,,
\label{eq:1Htetragonal2}
\end{align}
where we have defined
\begin{align}
F(k_{x},k_{y})&\equiv \cos k_{x}+\cos k_{y} 
+ \cos k_{x} \cos k_{y}.
\end{align}

We search for a condition that the model
has a finite parameter region in which a flat band emerges: 
If the two energy eigenvalues of the two-band Hamiltonian take a form, 
\begin{equation}
E=
\begin{cases}
\alpha \,+\, \beta\, F(k_{x},k_{y})\\
\gamma \,+\, \kappa\, F(k_{x},k_{y})
\end{cases},
\label{eq:idE}
\end{equation}
where $\alpha, \beta, \gamma, \kappa$ are real coefficients, 
we achieve a flat band when $\beta=0$ or $\kappa=0$ is satisfied.  
If we consider the characteristic polynomials for (\ref{eq:1Htetragonal}) and (\ref{eq:idE}), 
the former is
\begin{align}
&E^2 -\Big[(a+b)+2(d+e)F(k_{x},k_{y})\Big]E
\nonumber\\
&+ab+2(ae+bd) F(k_{x},k_{y})
+4de F(k_{x},k_{y})^2
\nonumber\\
&-4c^{2}[1+ F(k_{x},k_{y})]=0\,,
\end{align}
while the latter is
\begin{align}
&E^2 -\Big[(\alpha+\gamma)+(\beta+\kappa) F(k_{x},k_{y})\Big]E
\nonumber\\
&+\alpha\gamma+(\alpha\kappa+\beta\gamma) F(k_{x},k_{y})
+\beta\kappa F(k_{x},k_{y})^2=0\,.
\end{align}
For the two expressions to coincide with each other 
we have five equations,
\begin{align}
&a+b=\alpha + \gamma,
\nonumber\\
&2(d+e)= \beta+\kappa,
\nonumber\\
&ab-4c^{2}= \alpha\gamma,
\nonumber\\
&2ae+2bd-4c^2 = \alpha\kappa + \beta \gamma,
\nonumber\\
&4de=\beta\kappa.
\end{align}
When we eliminate $\alpha,\beta,\gamma,\kappa$, 
we obtain a single equation for $a,b,c,d,e$,
\begin{equation}
c^{2}\,=\, \frac{1}{2}(d-e)[2(d-e)-(a-b)],
\label{eq:fbc_tetragonal}
\end{equation}
which we call an ``auxiliary condition".  
Hereafter we impose this condition on the five parameters in Eq.(\ref{eq:hoptetragonal}), 
so that the model contains four independent parameters.  
We note that the auxiliary condition is a sufficient condition that
the model has a parameter region in which a flat band emerges.

By diagonalizing the Hamiltonian (\ref{eq:1Htetragonal}) with the auxiliary condition (\ref{eq:fbc_tetragonal}), 
we obtain the upper $(E^+)$ and lower $(E^-)$ bands as
\begin{align}
E^{\pm}&=
\begin{cases}
b+2d-2e + 2dF(k_{x},k_{y}) \\
a+2e-2d + 2eF(k_{x},k_{y})\,,
\end{cases}
\label{eq:Etetragonal}
\end{align}
where the parameter $c$ is eliminated through the auxiliary condition.
Obviously, the case of $d=0$ (i.e., $t_{\rm AA}^{(2)}=t_{\rm AA}^{(3)}=0$) 
or $e=0$ (i.e., $t_{\rm BB}^{(2)}=t_{\rm BB}^{(3)}=0$) yields a flat band.  
These energy eigenvalues are independent of choice of the sign in the auxiliary condition, 
$c=\pm\sqrt{(d-e)[d-e-(a-b)/2]}$. 
Although $c$ is eliminated, $|c|$ controls the band width, etc, indirectly 
through the auxiliary condition.

We note at this point that the energy of the flat band in the present model is nonzero in general, 
so that the flat band does not necessarily come 
from the rank reduction of the Hamiltonian as in Tasaki's two-band model. 
From Eq.~(\ref{eq:Etetragonal}), we find that the gap between the flat and dispersive bands is $b-a+2d$ for $e=0$ or $a-b+2e$ for $d=0$, which are obviously tunable.

Here, let us summarize the number of adjustable parameters in the model.
In the examples below, 
we set $|c|=1$ as a unit of energy and we are left with three independent parameters.
To obtain a flat band, we further set $d=0$ or $e=0$,
so that we are left with two adjustable parameters.
These two parameters enable us to control both the energy of the flat band and its gap from the dispersive band.
In the next two subsections, we also deal with a case of 
an extra condition imposed to classify the model, 
in which case we have only one adjustable parameter, with 
a Tasaki's model belonging to this.

\subsection{Non-rank-reduced type}
\begin{figure}[htbp]
\begin{center}
\includegraphics[width=0.47\textwidth]{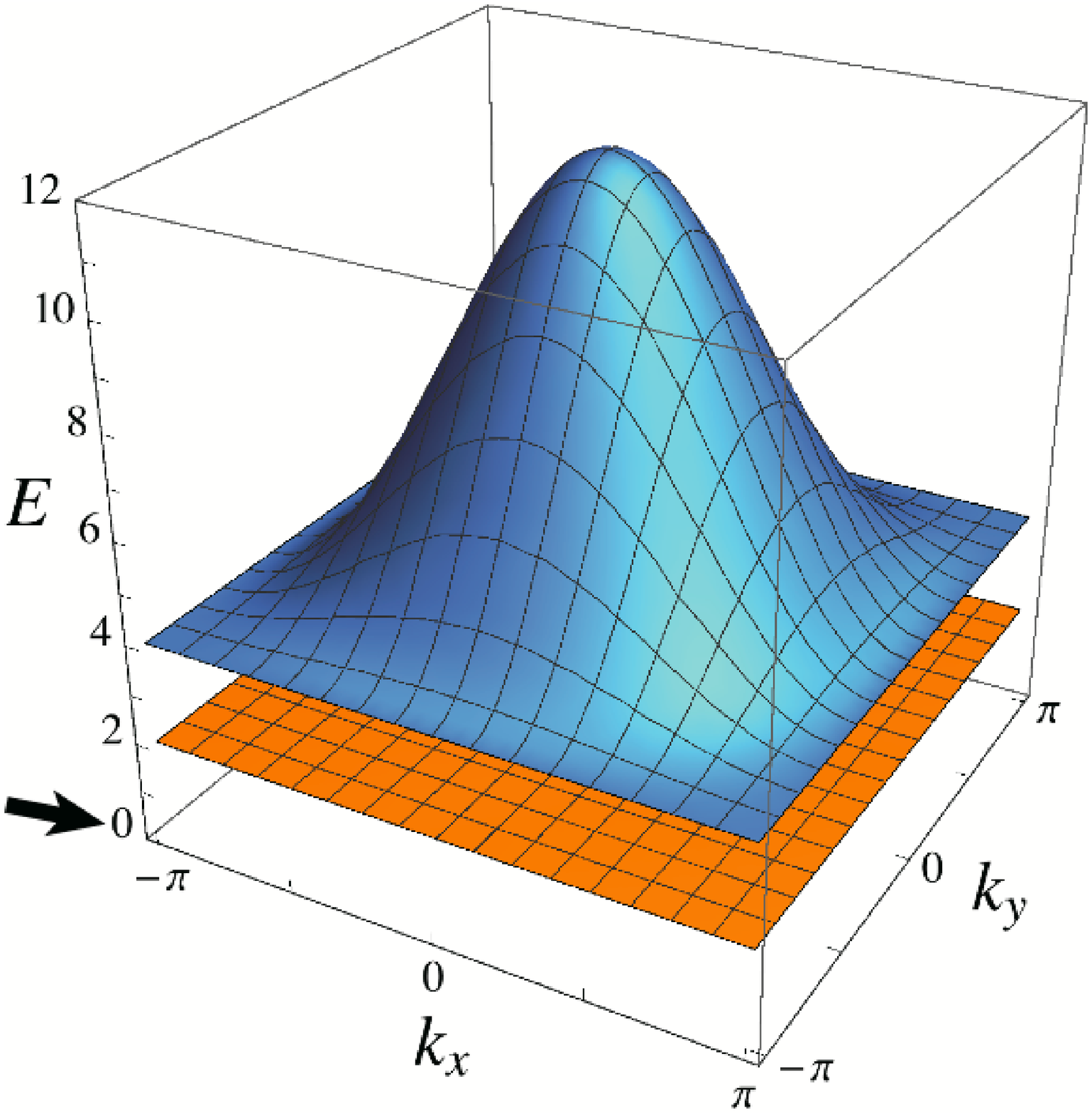}
\end{center}
\caption{An example of the band structures in the non-rank-reduced type for $(a,b,c,d,e)=(4,4,1,1,0)$.  An arrow in this and following figures indicates the 
position of zero energy.}
\label{fig:misumi01}
\end{figure}

For a general choice of the parameters the present model leads to a non-rank-reduced type. 
For example, if we consider a simple case of $a=b$ (no level offset for A and B sites), we end up with hopping parameters
\begin{align}
&{\rm A_{onsite}}\,\,:\,\, t_{\rm AA}^{(0)} =a, \quad
&{\rm B_{onsite}}\,\,:\,\, t_{\rm BB}^{(0)} =b=a,\nonumber\\
&{\rm I}\,\,:\,\,\quad\,t_{\rm AB}=c=\pm(d-e), \nonumber\\
&{\rm IIA}\,\,:\,\,t_{\rm AA}^{(2)} =d,\quad
&{\rm IIIA}\,\,:\,\, t_{\rm AA}^{(3)} = {d\over2},  \nonumber\\
&{\rm IIB}\,\,:\,\, t_{\rm BB}^{(2)} =e ,\quad
&{\rm IIIB}\,\, : \,\,t_{\rm BB}^{(3)} ={e \over2}, \nonumber\\
\label{eq:sites3-0}
\end{align}
where we have eliminated $b$ and $c$ by use of the auxiliary condition and $a=b$.
The eigenvalues are
\begin{align}
E^{\pm}=
\begin{cases}
a+2d-2e + 2dF(k_{x},k_{y}) \\
a+2e-2d + 2eF(k_{x},k_{y})\,.
\end{cases}
\end{align}
For $d=0$ or $e=0$,
a flat band emerges, which has a nonzero energy in general.
With fixing the unit of energy ($|c|=1$) and realizing a flat band (e.g. $e=0$) with the condition $a=b$, 
we are left with only one adjustable parameter ($a$) since $d$ is fixed by $c$ and $e$ via $c=\pm(d-e)$.
A typical example is illustrated in Fig.~\ref{fig:misumi01} for $(a,b,c,d,e)=(4,4,1,1,0)$.

\subsection{Rank-reduced type}
\label{subsec:rr}

\begin{figure}[htbp]
\begin{center}
\includegraphics[width=0.47\textwidth]{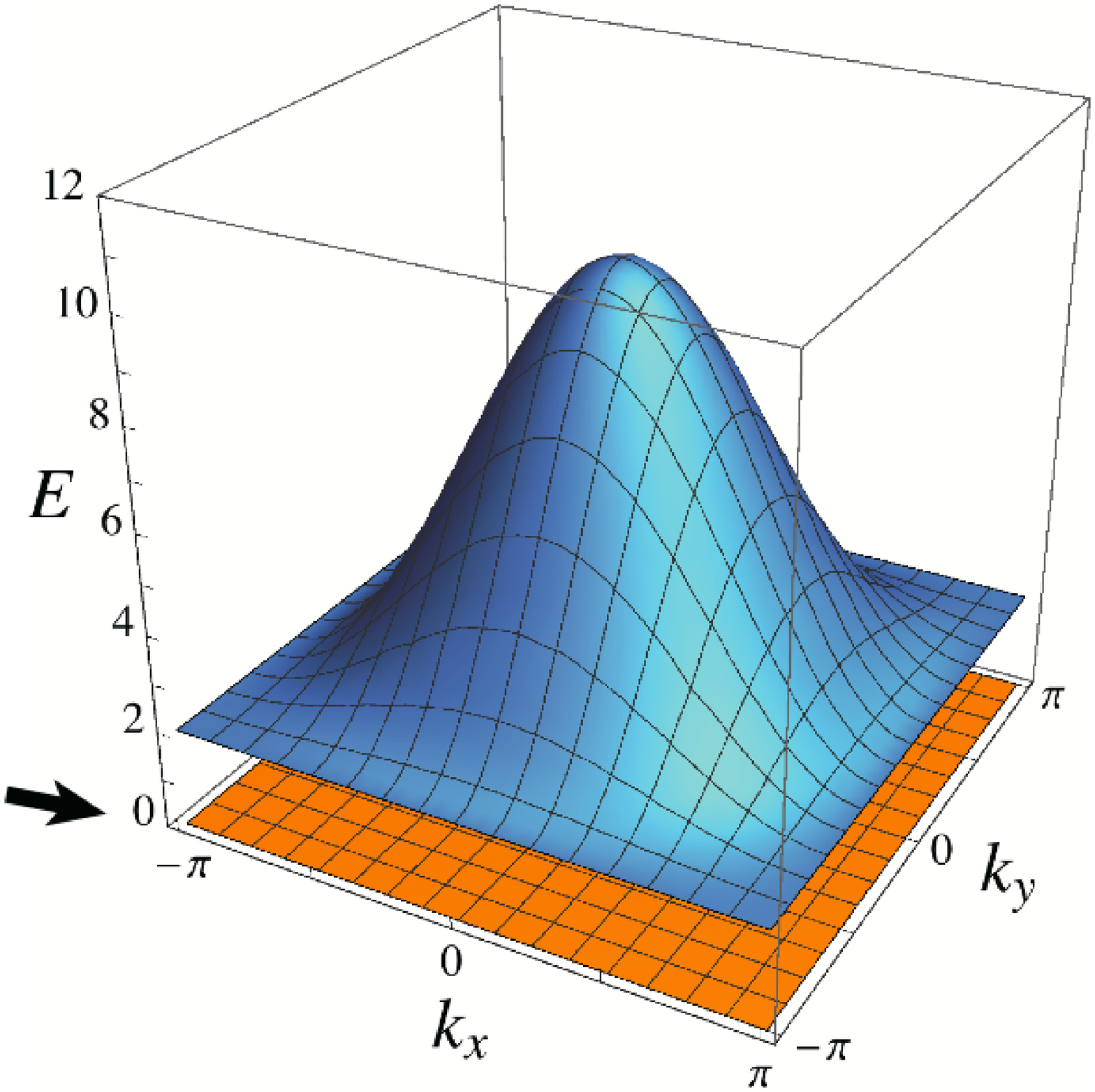}
\end{center}
\caption{An example of the band structures in the rank-reduced type 
for $(a,b,c,d,e)=(2,2,1,1,0)$.}
\label{fig:misumi_rr01}
\end{figure}

The conventional two-band flat-band models such as Tasaki's produce a flat band 
by reducing the rank of the Hamiltonian matrix from two to one.
We can realize this situation as well in the present model, Eq.(\ref{eq:hoptetragonal}) with Eq.(\ref{eq:fbc_tetragonal}), by e.g. imposing an extra condition that the wavenumber-{\it in}dependent part of the energy eigenvalues in Eq.(\ref{eq:Etetragonal}) be zero, 
i.e., $b +2d-2e \,=\,0$ 
or 
\begin{equation}
a+2e-2d \,=\,0 \,.
\label{eq:rrcon1}
\end{equation}
If we consider the latter case, 
the hopping parameters become
\begin{align}
&{\rm A_{onsite}}\,\,:\,\, t_{\rm AA}^{(0)} = a\,,\quad 
{\rm B_{onsite}}\,\,:\,\, t_{\rm BB}^{(0)} =b,\nonumber\\
&{\rm I}\,\,:\,\,\quad\,t_{\rm AB}=c = \pm {\sqrt{ab}\over{2}}, \nonumber\\
&{\rm IIA}\,\,:\,\,t_{\rm AA}^{(2)} =d={a+2e \over{2}},\nonumber\\
&{\rm IIIA}\,\,:\,\, t_{\rm AA}^{(3)} ={d\over{2}}= {a+2e \over4}, \nonumber\\
&{\rm IIB}\,\,:\,\, t_{\rm BB}^{(2)} =e ,\nonumber\\
&{\rm IIIB}\,\, : \,\, t_{\rm BB}^{(3)} ={e\over{2}}, \nonumber
\end{align}
where we have eliminated $c$ and $d$ with use of the auxiliary condition Eq.~(\ref{eq:fbc_tetragonal}) and the rank-reduction condition (\ref{eq:rrcon1}).
The energy eigenvalues in this case are
\begin{align}
E^{\pm}=
\begin{cases}
a+b + (a+2e)F(k_{x},k_{y}) \\
2eF(k_{x},k_{y})\,.
\end{cases}
\label{eq:bea1}
\end{align}
Here, $a+2e=0$ [i.e., $d=0$ from Eq.((\ref{eq:rrcon1})] or $e=0$ yield a flat band.  With $|c|=1$ and the flat-band condition (e.g. $e=0$) 
under the extra condition Eq.~(\ref{eq:rrcon1}), 
we are left with one adjustable parameter since $a$ and $b$ are related through $c=\pm \sqrt{ab}/2$.
A typical band structure is illustrated in Fig.~\ref{fig:misumi_rr01} for $(a,b,c,d,e)=(2,2,1,1,0)$ satisfying Eqs.(\ref{eq:fbc_tetragonal}),(\ref{eq:rrcon1}) and $e=0$, where the flat band is indeed located exactly at zero energy.

Another type of the rank-reduced flat-band models can be obtained 
if we impose that the wavenumber-independent part of the eigenvalues becomes zero simultaneously with the coefficients of the wavenumber-dependent part, i.e.,
\begin{align}
&b+2d-2e\to0 \,\,\;{\rm for}\;\,\, d=0
\nonumber\\
&\quad\quad\quad\quad\quad {\rm or}
\nonumber\\
&a+2e-2d\to0  \,\,\;{\rm for}\; \,\,e=0 \,. 
\label{eq:tetragonal-tasaki_c}
\end{align}
If this condition is satisfied, the relevant eigenvalues become zero. As a matter of fact, this case includes Tasaki's $(n,m)=(4,4)$ model \cite{Tasaki:1992} which is defined there in terms of three parameters denoted as 
$t,s,\nu (\in {\mathbb R})$. 
In our parametrization, Tasaki's model is given by
\begin{align}
&a=4t\nu^{2}-s,\quad b=t-4s\nu^{2}\,,
\nonumber\\
&c=\nu(t+s),\quad d=2\nu^2 t, \quad e=-2\nu^2 s\,,
\label{eq:tasaki_tetragonal1}
\end{align}
with $\nu\not= 0$ for the connectivity condition.
Namely we have
\begin{align}
&b+2d-2e\,=\,(1+4\nu^2)t, \\
&a+2e-2d\,=\,-(1+4\nu^2)s\,.
\end{align}
Since $d,e$ and $t,s$ are related as $d=2\nu^2 t$ and $e=-2\nu^2 s$, 
the condition Eq.~(\ref{eq:tetragonal-tasaki_c}) is satisfied,
i.e., setting $d=0$ or $e=0$ (equivalently $t=0$ or $s=0$) results in exactly zero-energy flat band.

\subsection{Connectivity condition}
In the known flat-band systems,  
Wannier orbitals cannot be orthogonalized and have an overlap with each other.
The condition for unorthogonalizable Wannier orbitals is called the ``connectivity condition" for the density matrix.  Intuitively, this condition is satisfied if arbitrary configurations of electrons on the lattice are realized just by moving a single hole.
 In Sec.~\ref{subsec:rr}, we have seen that the present model includes Tasaki's model, which is shown to satisfy the connectivity condition \cite{Tasaki:1998}. 
Even though the present class of models is broader than Tasaki's, 
the lattice structure and the pattern of hoppings in the two models are similar. In such bipartite models, the AB nearest hopping ($t_{\rm AB}=c$ here) connects a pair of lattice sites A and B, which is called an ``exchange bond". It was shown\cite{Tasaki:1998} that the connectivity condition is satisfied as long as the whole lattice is connected via the exchange bonds, or equivalently the hoppings for all the exchange bonds are nonzero. 
This situation is realized in the present model when $t_{\rm AB}=c \not=0$.
Therefore, the connectivity condition is considered to hold in the present model for $c\;(=1$ here$) \not=0$ with the Wannier orbitals unorthogonalizable.


\section{Modified models: crossing flat and dispersive bands}
\label{sec:tetragonald}

For the above flat-band model, we can prove that the flat and dispersive bands cannot overlap with each other, since the two energy eigenvalues in Eq.~(\ref{eq:Etetragonal}) cannot cross with each other for $d=0$ or $e=0$ 
with $c\not=0$:
Let us consider a case with $e=0$ in Eq.~(\ref{eq:Etetragonal}).
In order for the flat and dispersive bands to intersect,
the flat band with $E^{-}=a-2d$ needs to be between the top and bottom of the dispersive band $E^{+}=b+2d+2d(\cos k_{x}+\cos k_{y}+\cos k_{x}\cos k_{y})$, which means
\begin{equation}
0 \leq a-b-2d \leq 8d\,
\quad\quad\Rightarrow\quad\quad
0\leq -2c^{2}\leq 8d^{2}\,,
\end{equation}
where we have used the auxiliary condition (\ref{eq:fbc_tetragonal}) in 
the second expression.
The equation cannot be satisfied for $c\not=0$.
This implies that we have to introduce some 
modifications of the present flat-band models to realize crossing between the flat and dispersive bands. 
Here we consider two types of deformations.

\subsection{Type-I deformation}

In our first deformation (which we call Type-I) we deform the 
third-neighbor 
hoppings as
\begin{align}
&{\rm A_{onsite}}\,\,:\,\, t_{\rm AA}^{(0)}=a, \quad
&{\rm B_{onsite}}\,\,:\,\, t_{\rm BB}^{(0)}=b,\nonumber\\
&{\rm I}\,\,:\,\,\quad\, t_{\rm AB}=c, \nonumber\\
&{\rm IIA}\,\,:\,\,t_{\rm AA}^{(2)}=d,\quad
&{\rm IIIA}\,\,:\,\, t_{\rm AA}^{(3)}={d\over2}, \nonumber\\
&{\rm IIB}\,\,:\,\, t_{\rm BB}^{(2)}=e, \quad
&{\rm IIIB}\,\, : \,\, t_{\rm BB}^{(3)}={e(1-\epsilon)\over2}, \nonumber\\
\label{eq:sites_dm1}
\end{align}
where the third-neighbor 
hopping $t_{\rm BB}^{(3)}$ between B sites is reduced by $\epsilon$ with $0<\epsilon<2$. 
To retain the flat-band structure as much as possible,
we still impose the auxiliary condition Eq.(\ref{eq:fbc_tetragonal}) on the five parameters $a,b,c,d,e$ in Eq.~(\ref{eq:sites_dm1}).  
The Hamiltonian for (\ref{eq:sites_dm1}) becomes
\begin{align}
&{\mathcal H}_{\rm AA}  = a+  2d(\cos k_{x}+\cos k_{y} + \cos k_{x} \cos k_{y})\,,
\nonumber\\
&{\mathcal H}_{\rm AB} = {\mathcal H}_{\rm BA} = 4c\cos{k_{x}\over{2}}\cos {k_{y}\over{2}} \,,
\nonumber\\
&{\mathcal H}_{\rm BB} = b+ 2e[\cos k_{x}+\cos k_{y} +(1-\epsilon)\cos k_{x} \cos k_{y}]\,.
\end{align}
While the expression for the energy eigenvalues obtained by diagonalizing this matrix is quite complicated (which can be obtained with {\it Mathematica}), 
we can numerically see that a nearly flat band emerges for $d=0$ as we shall see.
In the following examples, we again set $|c|=1$ as a unit of energy.

\begin{figure}[htbp]
\begin{center}
\includegraphics[width=0.45\textwidth]{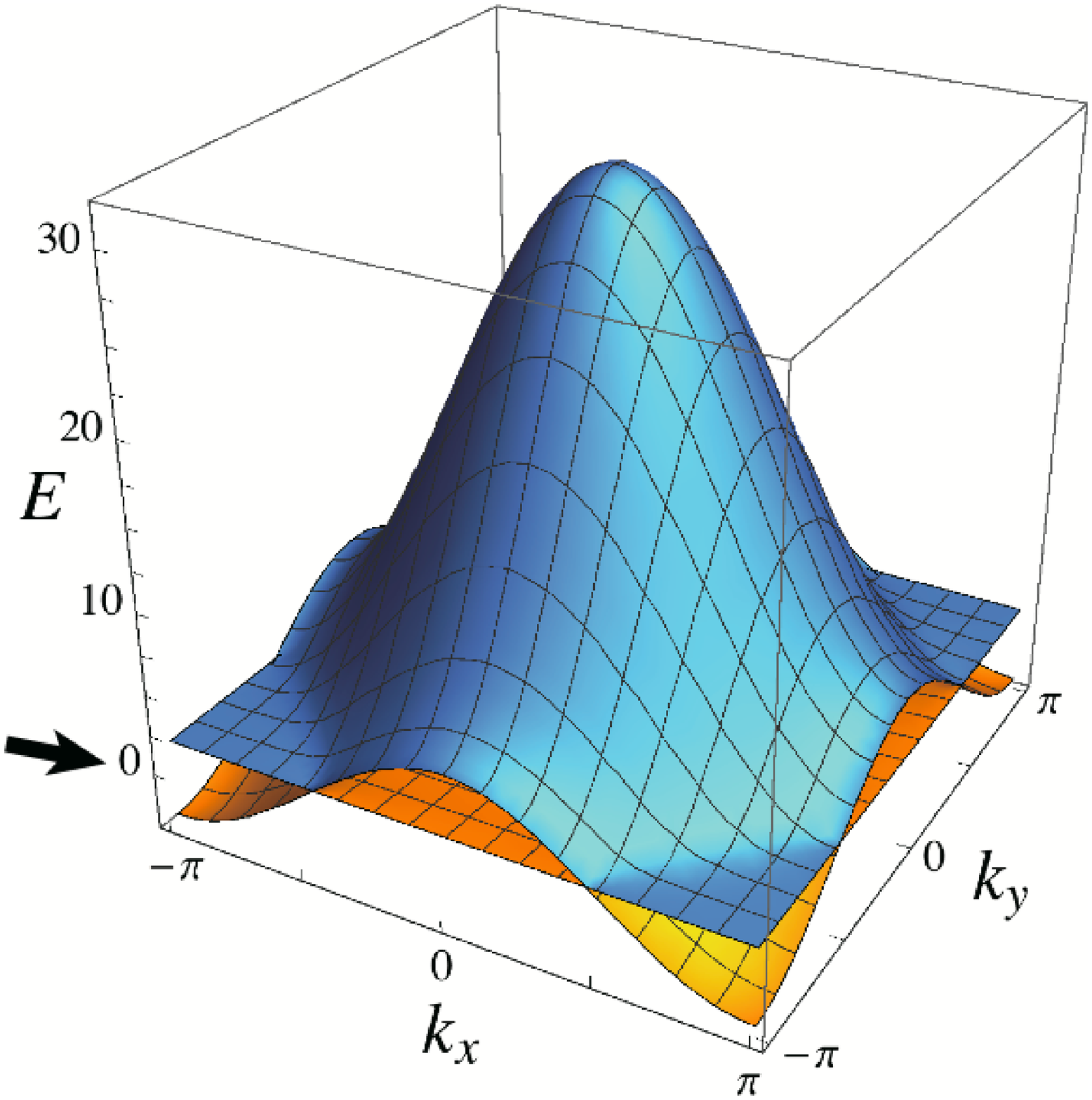}
\includegraphics[width=0.40\textwidth]{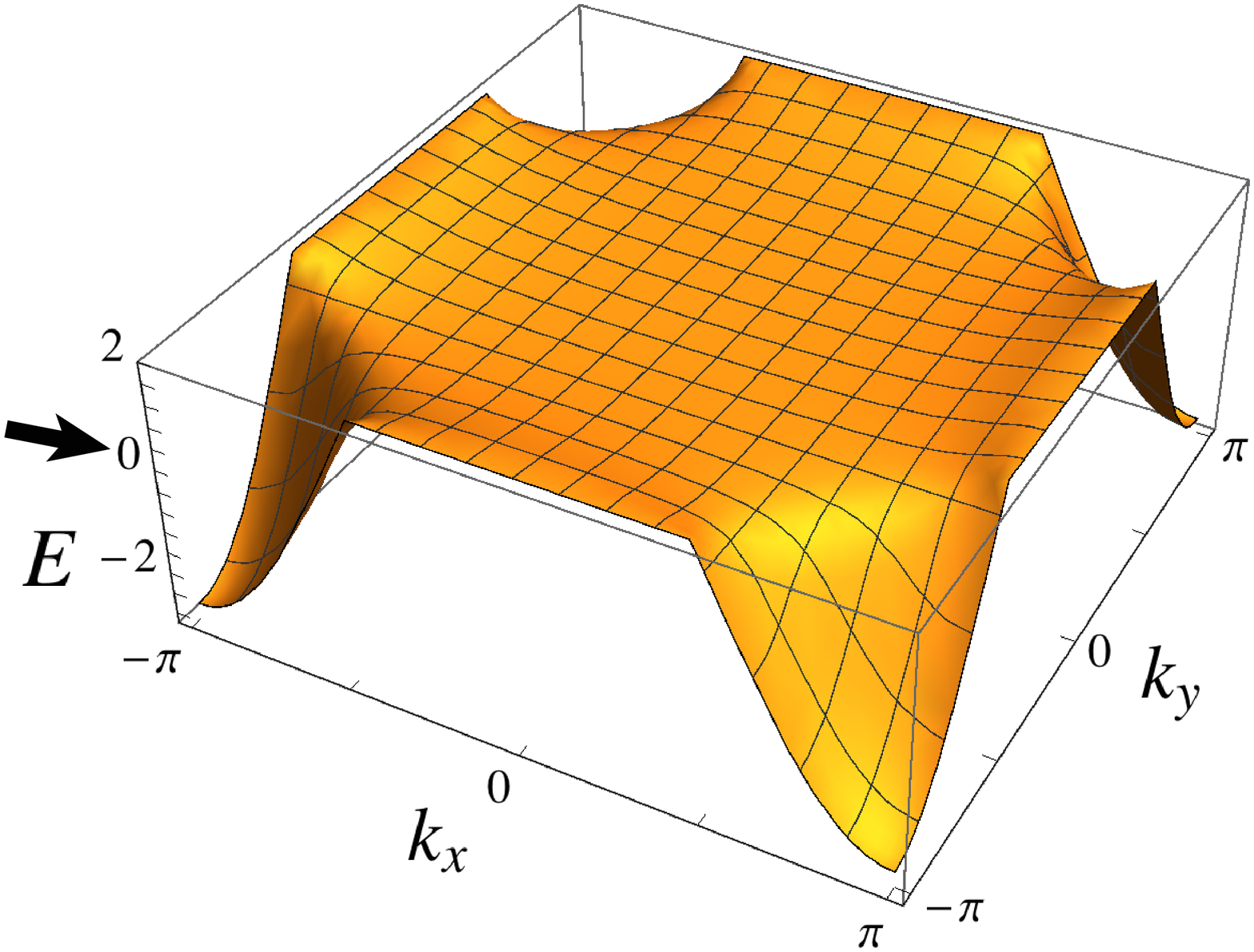} 
\end{center}
\caption{
Top: The band structure in the Type-I deformed tetragonal flat-band model for $(a,b,c,d,e)=(2,10,1,0,2+\sqrt{5})$ with $\epsilon=1/2$. Bottom: The lower band is separately depicted.}
\label{fig:misumidf01}
\end{figure}

\begin{figure}[htbp]
\begin{center}
\includegraphics[width=0.43\textwidth]{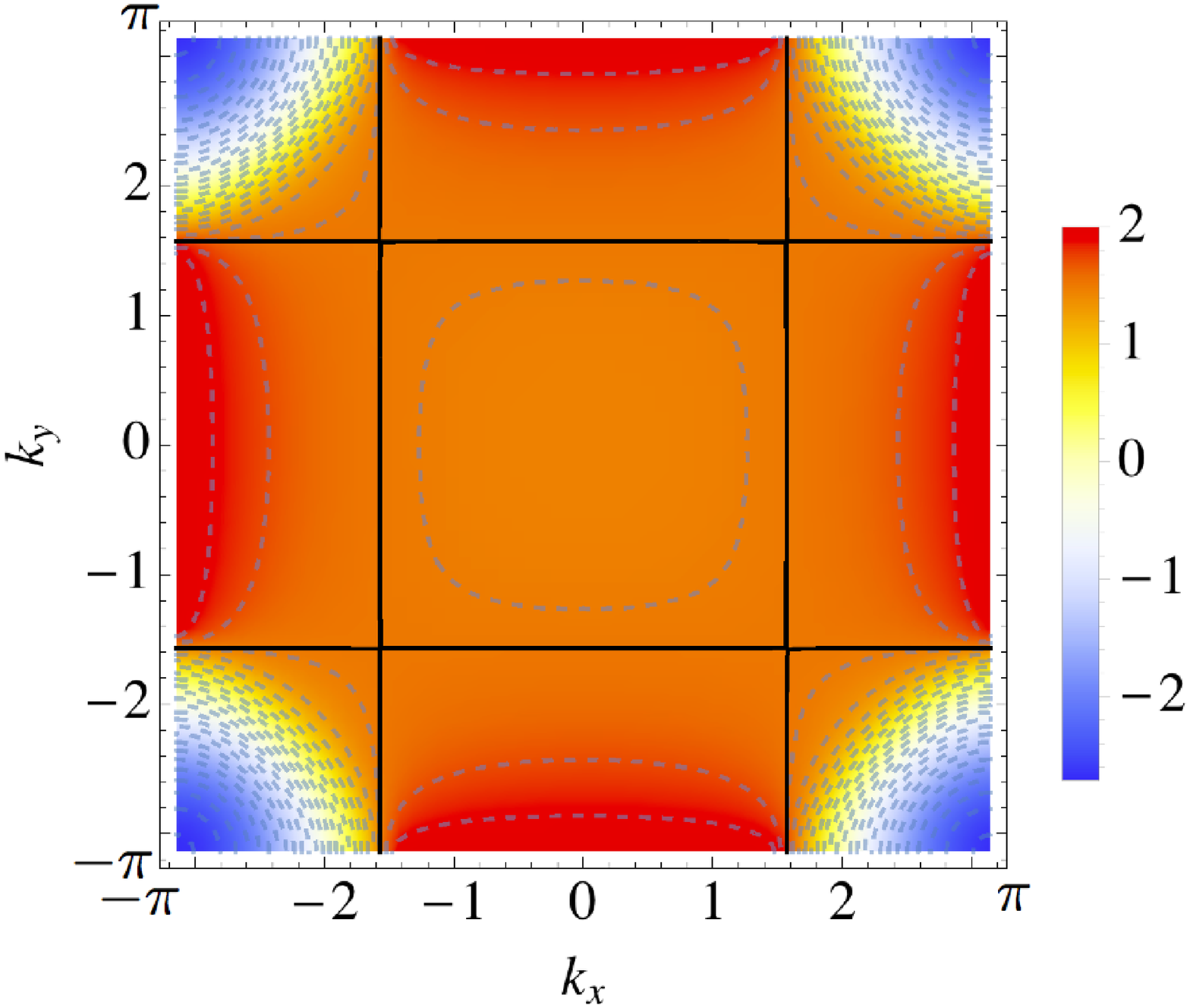} 
\end{center}
\caption{Contour plot of the lower band for the same parameter set as in 
Fig.\ref{fig:misumidf01}. Fermi surface for the half-filled lower band at $\mu=1.528$ is depicted as bold lines.  }
\label{fig:s-13nu3C}
\end{figure}

A typical example is displayed in Fig.~\ref{fig:misumidf01} for 
$(a,b,c,d,e)=(2,10,1,0,2+\sqrt{5})$ with $\epsilon=1/2$, which satisfies the auxiliary condition Eq.(\ref{eq:fbc_tetragonal}).
We can see that there is a nearly-flat band 
that intersects a dispersive one. 
To be more precise, each of the lower and upper bands is composed of nearly-flat and dispersive parts. 
For fixed $a=2,c=1,d=0,\epsilon=1/2$, we can adjust the flatness 
by varying $b$: 
the larger the $b$, the stronger the flatness.  
We also depict the contour plot of the lower band in Fig.~\ref{fig:s-13nu3C}, 
where we indicate the half-filled Fermi surface as bold lines, whose
chemical potential $\mu=1.528$ is determined numerically.  
Here ``half-filling" means a situation that the lower band is half-filled, 
where the total filing factor is $N_{\rm e}/N_{\rm s}=1/4$ with the numbers of electrons  and sites being $N_{\rm e}$ 
and $N_{\rm s}$ respectively.  The Fermi surface happens to have 
nesting vectors $(0, \pm\pi), (\pm\pi, 0)$ for this choice of the 
Fermi energy.

\begin{figure}[htbp]
\begin{center}
\includegraphics[width=0.45\textwidth]{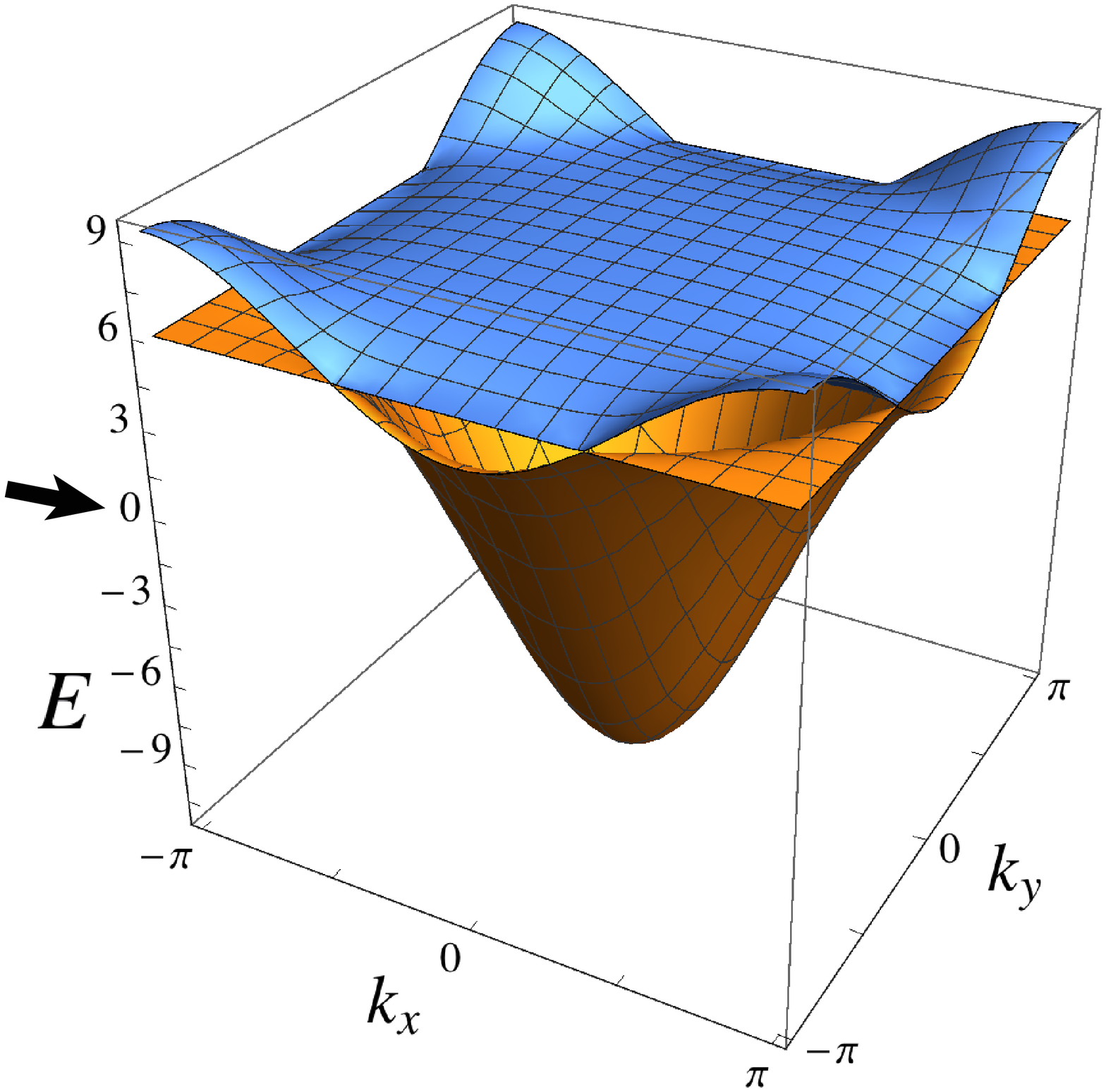}
\end{center}
\caption{The band structure in the Type-I deformed flat-band model for 
a negative value of $e$ with $(a,b,c,d,e)=(6,2,1,0,-1-\sqrt{2})$ and $\epsilon=1/2$.}
\label{fig:misumidf02}
\end{figure}

We can also construct an inverted band structure, as exemplified in 
Fig.~\ref{fig:misumidf02} for $(a,b,c,d,e)=(6,2,1,0,-1-\sqrt{2})$ and 
$\epsilon=1/2$ with the auxiliary condition Eq.(\ref{eq:fbc_tetragonal}) again satisfied.  
We note that a negative $e$ leads to an inverted band, 
with $c$ still positive.

We can note that the lower and upper bands in the Type-I deformed model 
touch with each other at two points on each edge of the Brillouin zone ($k_{x}=\pm\pi$ or $k_{y}=\pm\pi$) 
as seen in Fig.~\ref{fig:misumidf01}. 
By comparing the energy eigenvalues of the lower and upper bands,
we find that the touching occurs as long as $|(a-b+2e)/(2e\epsilon)| < 1$ is satisfied.

\subsection{Type-II deformation}

Alternatively, we can modify the second-neighbor hopping on B sites, $t^{(2)}_{\rm BB}$, to have another deformation (which we call Type-II) as
\begin{align}
&{\rm A_{onsite}}\,\,:\,\, t_{\rm AA}^{(0)}=a,\quad
&{\rm B_{onsite}}\,\,:\,\, t_{\rm BB}^{(0)}=b,\nonumber\\
&{\rm I}\,\,:\,\,\quad\, t_{\rm AB}=c, \nonumber\\
&{\rm IIA}\,\,:\,\,t_{\rm AA}^{(2)}=d,\quad
&{\rm IIIA}\,\,:\,\, t_{\rm AA}^{(3)}={d\over2}, \nonumber\\
&{\rm IIB}\,\,:\,\, t_{\rm BB}^{(2)}=e(1-\epsilon),\quad
&{\rm IIIB}\,\, : \,\, t_{\rm BB}^{(3)}={e\over2}, \nonumber\\
\label{eq:sites_dm2}
\end{align}
where we again preserve the auxiliary condition Eq.(\ref{eq:fbc_tetragonal}).  
The Hamiltonian in $k$-space now becomes
\begin{align}
&{\mathcal H}_{\rm AA}  = a+  2d(\cos k_{x}+\cos k_{y} + \cos k_{x} \cos k_{y})\,,
\nonumber\\
&{\mathcal H}_{\rm AB} = {\mathcal H}_{\rm BA} = 4c\cos{k_{x}\over{2}}\cos {k_{y}\over{2}} \,,
\nonumber\\
&{\mathcal H}_{\rm BB} = b+ 2e[(1-\epsilon)(\cos k_{x}+\cos k_{y}) +\cos k_{x} \cos k_{y}]\,.
\end{align}

\begin{figure}[htbp]
\begin{center}
\includegraphics[width=0.45\textwidth]{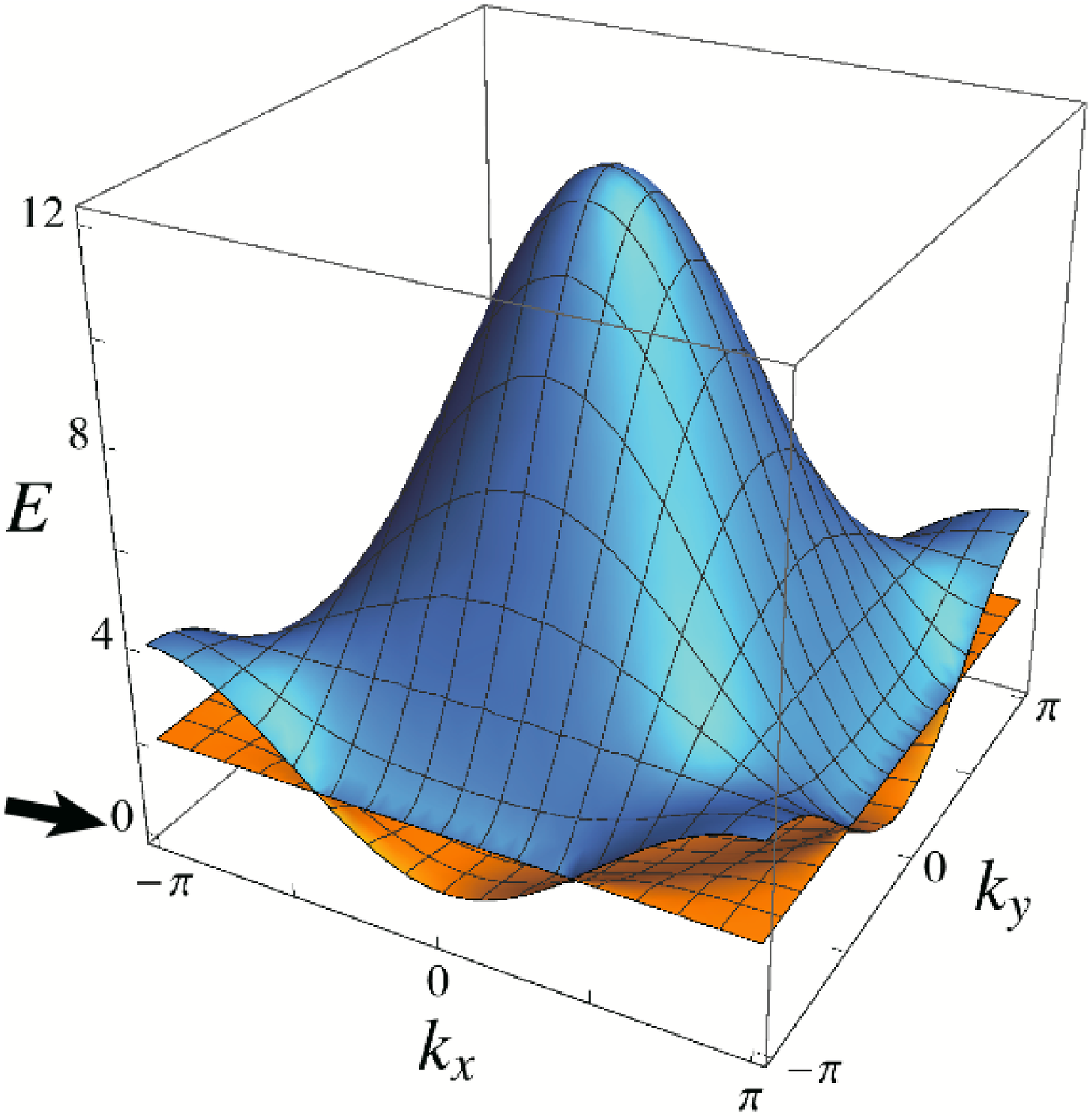}
\includegraphics[width=0.32\textwidth]{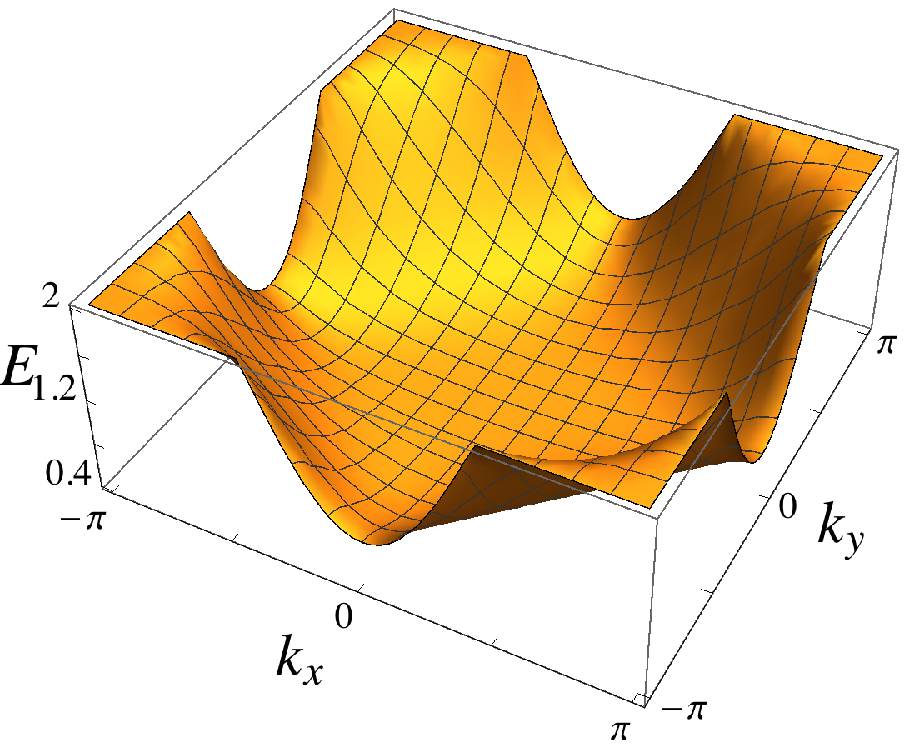} 
\end{center}
\caption{Top: The band structure in the Type-II deformed tetragonal flat-band model for $(a,b,c,d,e)=(2,4,1,0,(1+\sqrt{5})/2)$  with $\epsilon=1/2$. 
Bottom: The lower band is separately depicted.}
\label{fig:misumidf03}
\end{figure}

\begin{figure}[htbp]
\begin{center}
\includegraphics[width=0.43\textwidth]{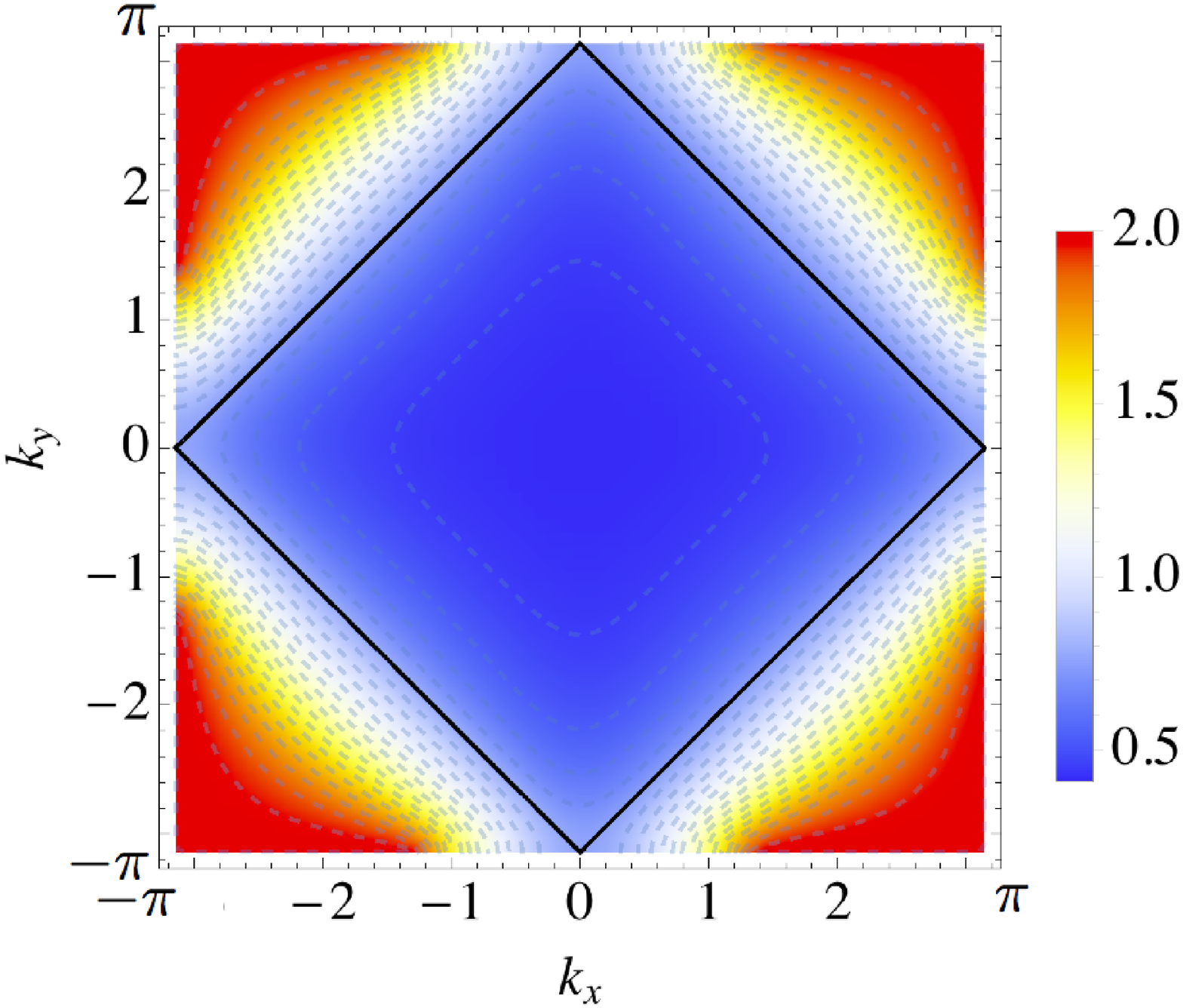} 
\end{center}
\caption{Contour plot of the lower band in the Type-II deformed tetragonal flat-band model for the same parameter set as in Fig.\ref{fig:misumidf03}. 
Fermi surface for a chemical potential $\mu=0.764$ 
is depicted as a bold line.}
\label{fig:fig034}
\end{figure}

In this case a nearly flat band emerges for $d=0$. 
A typical example is displayed  in Fig.~\ref{fig:misumidf03} for 
$(a,b,c,d,e)=(2,4,1,0,(1+\sqrt{5})/2)$ with $\epsilon=1/2$.   
Although a nearly, and partly, flat band intersects a dispersive one in this case too, 
the Fermi surface for the half-filled lower band with the chemical potential $\mu=0.764$, 
shown in Fig.~\ref{fig:fig034}, is diamond-shaped 
[with nesting vectors $(\pm\pi, \pm\pi)$] and 
differs from those in Type-I deformation.  
Here ``half-filling" again means $N_{\rm e}/N_{\rm s}=1/4$.

\subsection{Possibility of superconductivity}

At this point let us discuss one implication of the band structures obtained here. As mentioned in the Introduction, it has been proposed that, when flat band and dispersive band are intersecting, the virtual pair scatterings between the two bands can lead to relatively high-$T_{C}$ superconductivity \cite{Kuroki:2005kha,Takayoshi:2013tkw,Tovmasyan:2013tnh,Kobayashi:2016koy}.  
In particular, quasi-one-dimensional systems with the flat-dispersive band crossing were shown to exhibit a relatively high-$T_{C}$ superconductivity, 
with the fluctuation-exchange approximation (FLEX) \cite{Kuroki:2005kha} and with the density-matrix renormalization group (DMRG)\cite{Kobayashi:2016koy}.  
In these quasi-1D models, however, the dispersive band has a 
$k$-linear dispersion around the band crossing.  This makes the 
density of dispersive states finite in 1D cases around the flat-band energy.  
When we go over to 2D models, however, Lieb's model has  a 
dispersion linear in $k_x$ and $k_y$ (i.e., a ``Dirac cone") around the band crossing, which implies the density of dispersive states around the flat-band energy 
vanishes since we are now in 2D.  Then what we want, for a nonzero density 
of states, is a flat band piercing a {\it parabolic} band (or more 
generally a cosine band).  
The deformed  flat-band model we have constructed in Section III does realize such an intersection between the (nearly) flat and dispersive bands. 
If we consider the case in which the Fermi energy is close to, 
but away from, the flat part of the bands,
then the virtual pair scatterings between flat and dispersive parts 
may favor superconductivity.  In the 1D case it is shown\cite{Kuroki:2005kha} 
that such superconductivity persists even when the flat band is warped to some extent, since the virtual pair hopping can still work.  
So it will be an intriguing question to ask whether the superconductivity in 2D such as $d$-wave pairing could be enhanced in the present model.  

\begin{figure}[t]
\begin{center}
\includegraphics[width=0.42\textwidth]{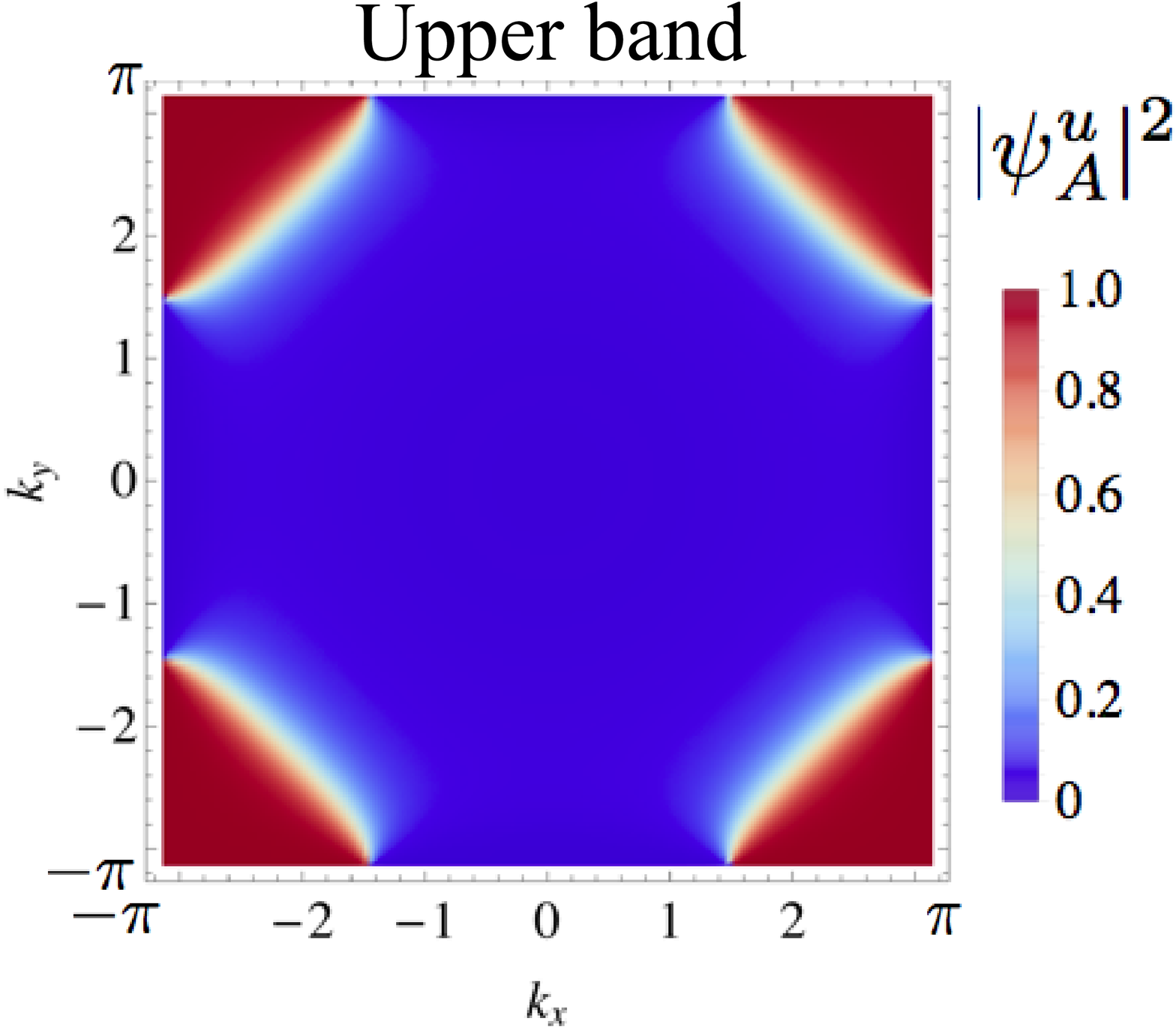} 
\includegraphics[width=0.42\textwidth]{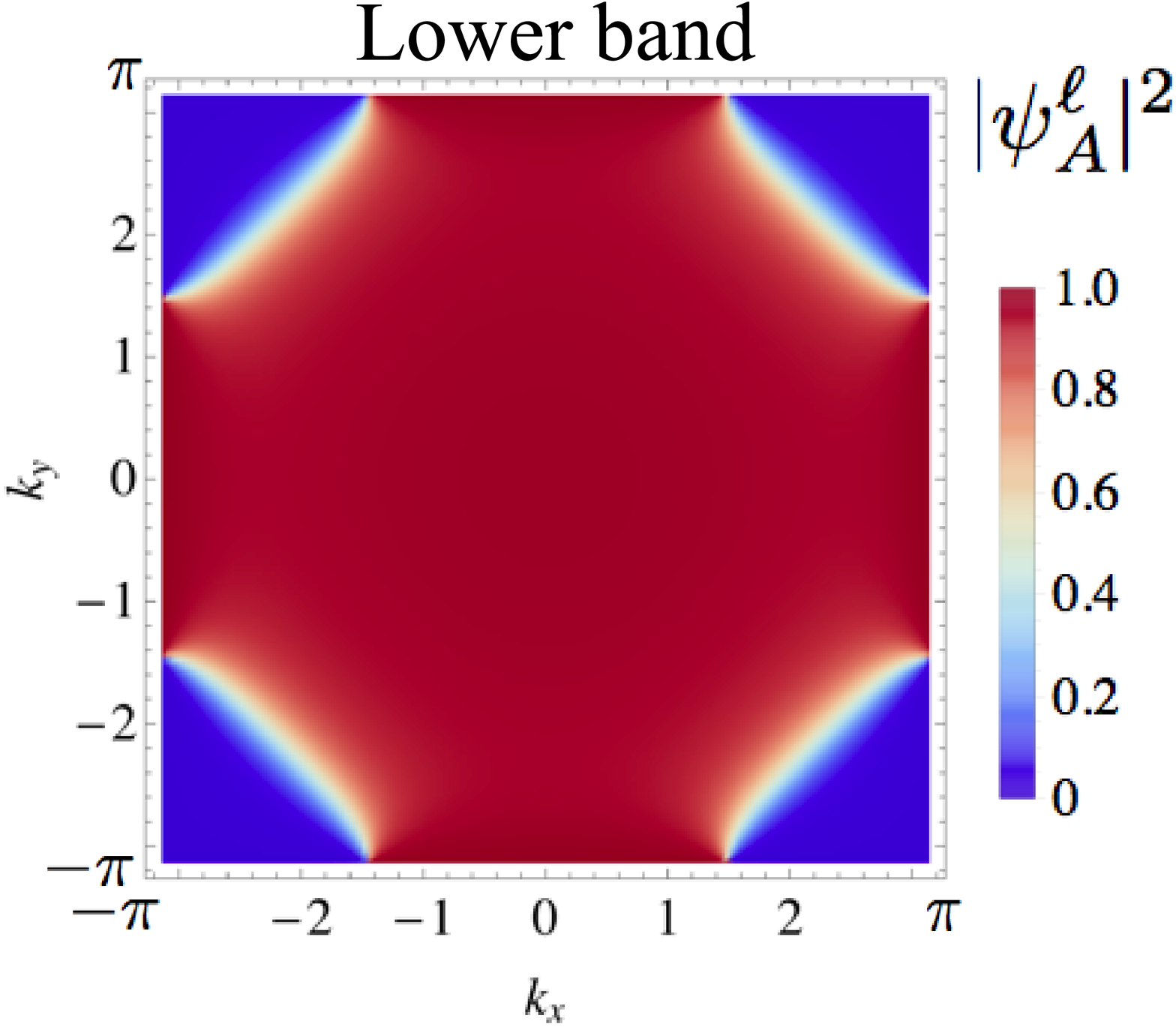} 
\end{center}
\caption{For Type-I deformed model, 
a wavefunction component for A orbital $|\psi_{A}|^2$ with $|\psi_{B}|^2=1-|\psi_{A}|^2 $ is depicted for the upper (top) and lower (bottom) bands in Fig.~\ref{fig:misumidf01}. Red (blue) regions represent A(B)-dominated ones, while 
white regions AB-mixed ones.
}
\label{fig:wf1}
\end{figure}

\begin{figure}[t]
\begin{center}
\includegraphics[width=0.42\textwidth]{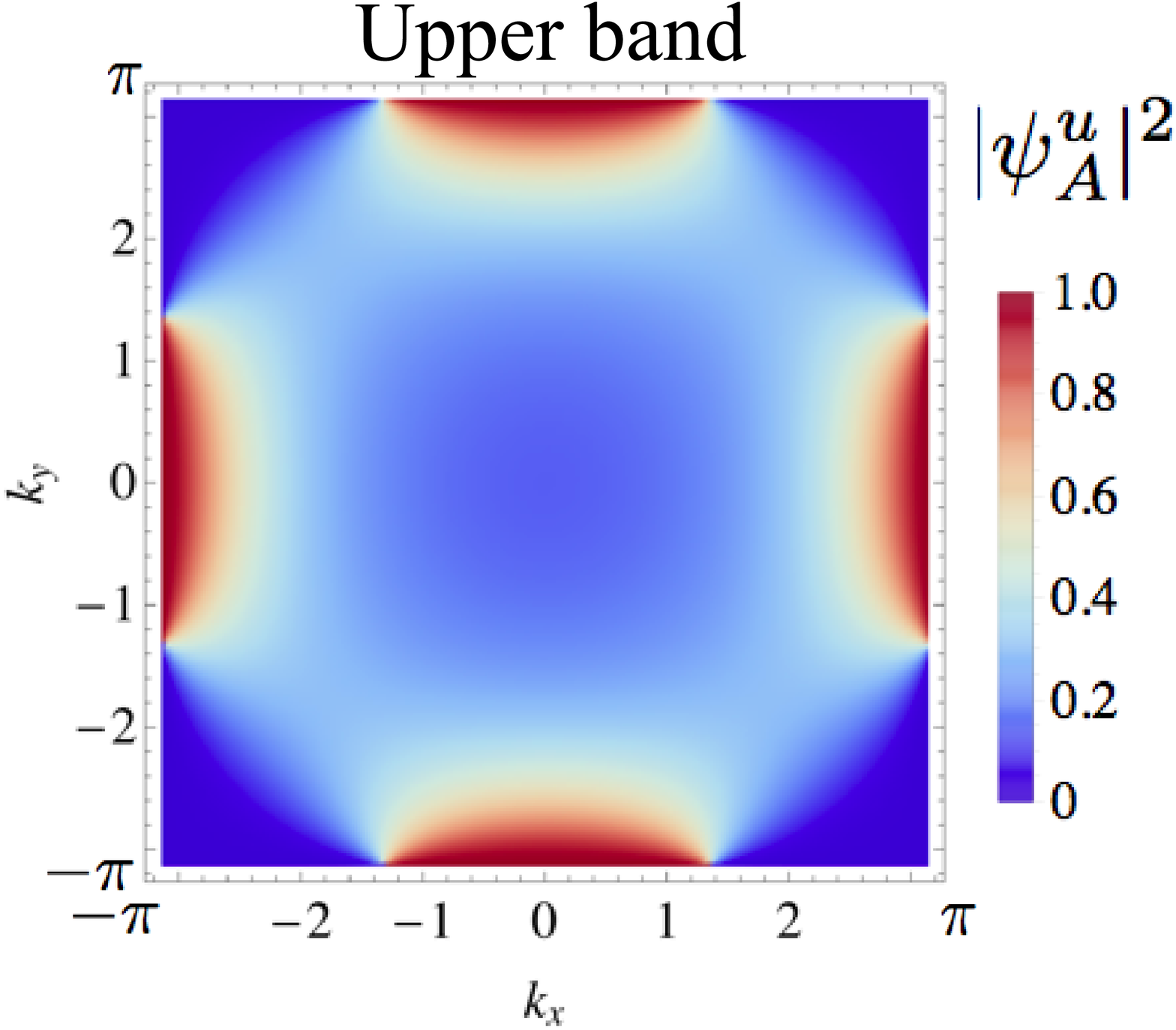} 
\includegraphics[width=0.42\textwidth]{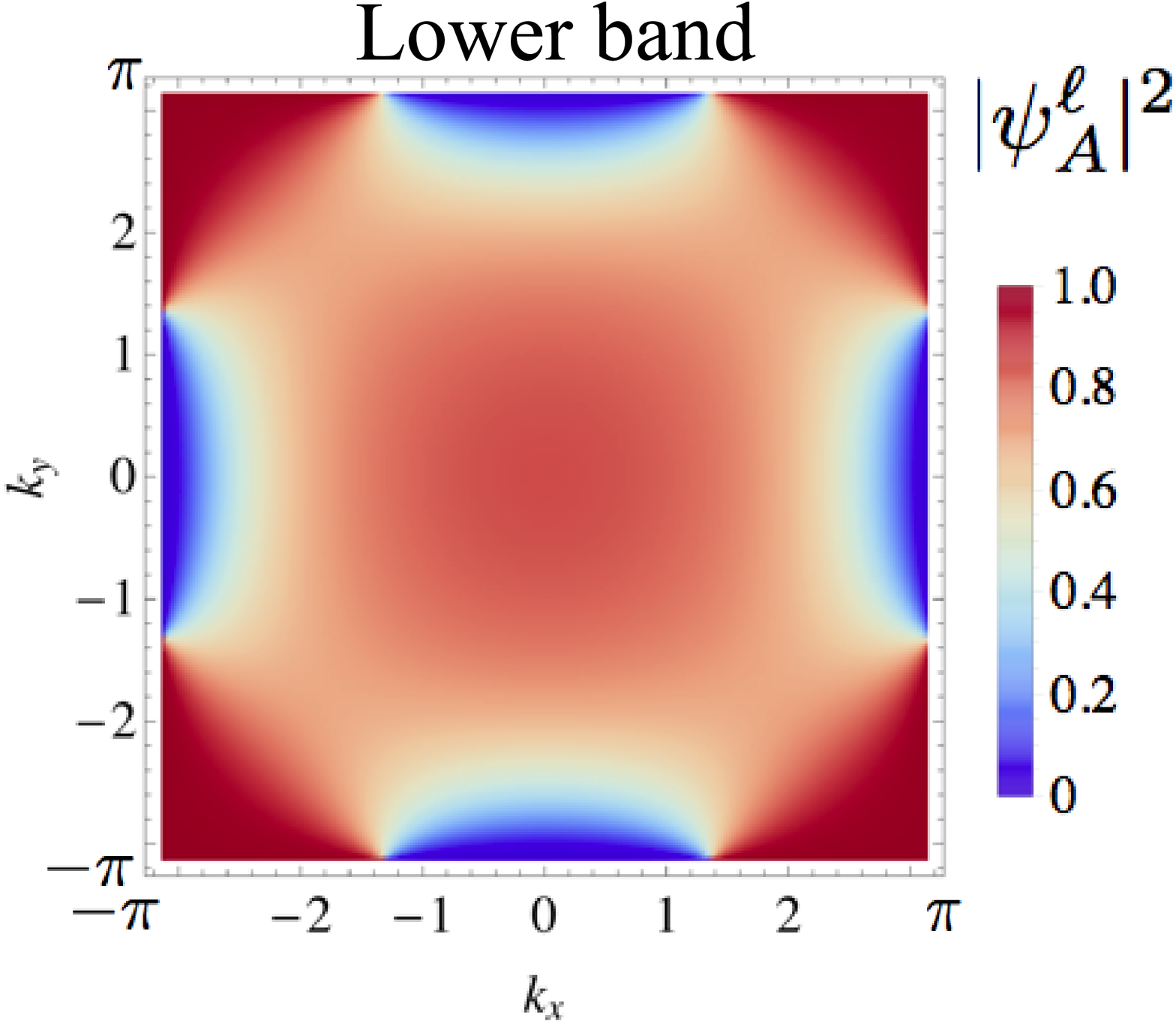}
\end{center}
\caption{For Type-II deformed model, 
a wavefunction component for A orbital $|\psi_{A}|^2$ with $|\psi_{B}|^2=1-|\psi_{A}|^2 $ is depicted for the upper (top) and lower (bottom) bands in Fig.~\ref{fig:misumidf03}. Red (blue) regions represent A(B)-dominated ones, while 
white regions AB-mixed ones.}
\label{fig:wf2}
\end{figure}

The spin-fluctuation mediated pairing between flat and dispersive bands is enhanced 
when the virtual pair scattering occurs on the same sublattice (A or B) 
as suggested from the two-band FLEX \cite{Kuroki:2017pcm}.
It means that it is better to avoid a case in which the A-component is dominant on the flat band while 
the B-component is dominant on the dispersive band (or vice versa).  
So let us now look into the wavefunction character in Fig.~\ref{fig:wf1} for the upper and lower bands in the Type-I deformed model (Fig.~\ref{fig:misumidf01}), and in Fig.~\ref{fig:wf2} for those in the Type-II deformed model (Fig.~\ref{fig:misumidf03}). We here depict the A-component weight $|\psi_{A}^{u}({\bf k})|^{2}$ for the wavefunctions of the upper band and $|\psi_{A}^{\ell}({\bf k})|^{2}$ for the lower band. The B-component weights are $|\psi_{B}^{u}({\bf k})|^{2}=1-|\psi_{A}^{u}({\bf k})|^{2}$ and $|\psi_{B}^{\ell}({\bf k})|^{2}=1-|\psi_{A}^{\ell}({\bf k})|^{2}$.

In Fig.~\ref{fig:wf1} for the Type-I deformed model, we find that the A-component is dominant in the flat parts while the B-component is dominant in the dispersive parts.
However, the A-component remains in the dispersive parts close to the 
energy regions of the flat parts: 
Typically, ${\bf k}=(2\pi/3,2\pi/3)$ belongs to dispersive parts both in 
upper and lower bands but sits close to the flat parts (yellow region in Fig.~\ref{fig:s-13nu3C}), 
where we have $|\psi_{A}^{u}(2\pi/3,2\pi/3)|^{2}=0.36$ and $|\psi_{A}^{\ell}(2\pi/3,2\pi/3)|^{2}=0.64$.  
This exemplifies that the dispersive part can have significant components that are dominant in the flat parts.  
Thus, when the Fermi energy is away from, but close to the flat parts, a Type-I deformed model can accommodate pair hoppings between flat and dispersive parts.

In the Type-II deformed model, this occurs even more conspicuously as seen in 
Fig.~\ref{fig:wf2}, where we can see that the A-component dominates both the flat and dispersive parts of the lower band: For example, we find that ${\bf k}=0$ belonging to the flat part of the lower band (as seen from Fig.~\ref{fig:misumidf03}) 
has $|\psi_{A}^{\ell}({\bf k}=0)|^2 =0.86$ while ${\bf k}=(2\pi/3,2\pi/3)$ 
in the dispersive part has $|\psi_{A}^{\ell}(2\pi/3,2\pi/3)|^2 =0.76$.  
The upper band is mainly dominated by the B-component, but 
the A-component in the dispersive part is appreciable around the region 
intersecting with the lower band (e.g., $|\psi_{A}^{u}(8\pi/9, \pi/3)|^{2}=0.59$).
These suggest that the Type-II deformed model can lead to even more 
significant pair hoppings between the flat and dispersive bands.  

As for the A and B sublattices, we can make another comment, 
on the relation with a preceding research:
Imada {\it et al}. considered a flat-band model in which 
each lattice site having usual hopping is accompanied by 
extra orbitals with no hopping to adjacent sites.  
This may also be regarded as a three-layer model with intersite 
hoppings existing only on one layer.  
They argue that a high-$T_{C}$ superconductivity could be realized when the Fermi energy is set right on the flat band\cite{Imada:2000iko}.
In the present models, on the other hand, 
we consider single-layer lattice systems with distant hoppings, 
and also we have in mind the flat band as intermediate states for virtual pair scatterings for $E_F$ sitting away from the band. 
Still, the relation between these models will be interesting.

\section{Flat-band model on honeycomb lattice}
\label{sec:honey}

\begin{figure}[htbp]
\begin{center}
\includegraphics[width=0.47\textwidth]{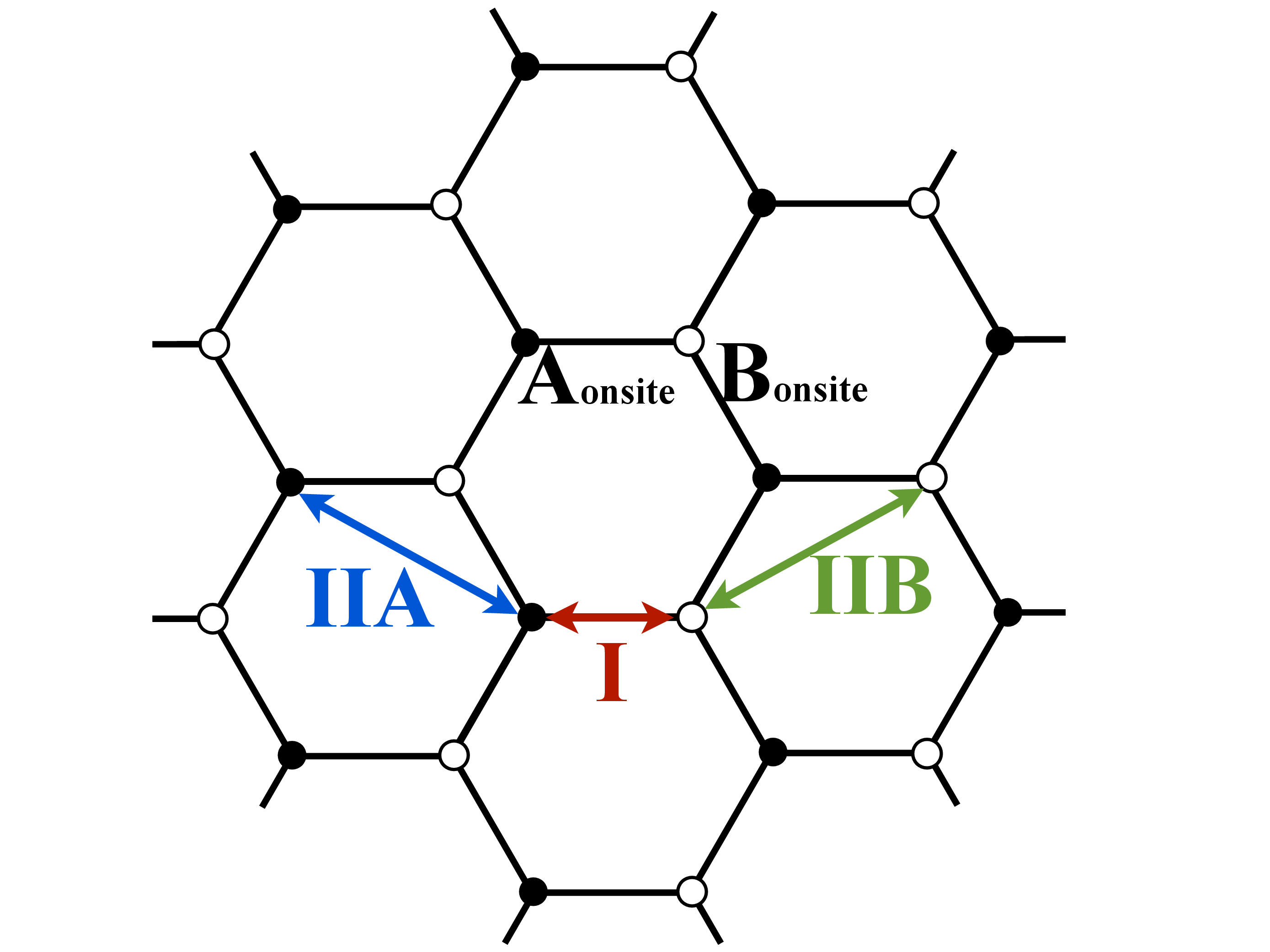}
\end{center}
\caption{Hopping parameters for the honeycomb flat-band model. 
Solid and open circles represent A and B sublattice sites, respectively.}
\label{fig:sites_honey}
\end{figure}

In order to show that the present scheme is not restricted to tetragonal (cubic in general dimensions) lattices,
let us now consider a tight-binding model on honeycomb lattice  
(Fig.~\ref{fig:sites_honey}).  
Here we only consider the nearest(I) and second (II) neighbor hoppings as 
\begin{align}
&{\rm A_{onsite}}\,\,:\,\, t_{\rm AA}^{(0)} = a,\quad
{\rm B_{onsite}}\,\,:\,\, t_{\rm BB}^{(0)} = b, \nonumber\\
&{\rm I}\,\,:\,\,\quad\,t_{\rm AB} = c, \nonumber\\
&{\rm IIA}\,\,:\,\,t_{\rm AA}^{(2)} = d,\quad
{\rm IIB}\,\,:\,\, t_{\rm BB}^{(2)} = e,\nonumber\\
\label{eq:hophoney}
\end{align}
with five free parameters $a,b,c,d,e \in {\mathbb R}$.
The Hamiltonian is
\begin{align}
{\mathcal H}=
\begin{pmatrix}
{\mathcal H}_{\rm AA}  & {\mathcal H}_{\rm AB} \\
{\mathcal H}_{\rm BA} & {\mathcal H}_{\rm BB}
\end{pmatrix}\,,
\label{eq:1Hhoney}
\end{align}
with
\begin{align}
&{\mathcal H}_{\rm AA}  = a\,+\,  d\,F^{\rm honeycomb}(k_{x},k_{y})\,,
\nonumber\\
&{\mathcal H}_{\rm AB} = {\mathcal H}_{\rm BA}^{*} = c\left[e^{i{k_{x}\over{\sqrt{3}}}} 
+ e^{i\left( - {\sqrt{3}k_{x}\over{6}}+{k_{y}\over{2}}\right)}+ e^{i\left( - {\sqrt{3}k_{x}\over{6}}-{k_{y}\over{2}}\right)}\right],
\nonumber\\
&{\mathcal H}_{\rm BB} = b\,+\,  e\,F^{\rm honeycomb}(k_{x},k_{y})\,.
\label{eq:1Hhoney2}
\end{align}
Here,  $k_{x}$ and $k_{y}$ are wavenumbers on a Cartesian coordinates, 
and we have defined
\begin{align}
F^{\rm honeycomb}(k_{x},k_{y}) \equiv 4\cos {\sqrt{3}k_{x}\over{2}}\cos {k_{y}\over{2}} +  2\cos k_{y}.
\end{align}

A procedure to obtain the auxiliary condition, i.e., a condition that a model
has finite parameter space in which a flat band can emerge, is parallel to that for the tetragonal model: 
If energy eigenvalues of the Hamiltonians (\ref{eq:1Hhoney})(\ref{eq:1Hhoney2}) take a form
\begin{equation}
E^{\pm}=
\begin{cases}
\alpha \,+\, \beta\, F^{\rm honeycomb}(k_{x},k_{y})\\
\gamma\,+\, \kappa\, F^{\rm honeycomb}(k_{x},k_{y})\,,
\end{cases}
\label{eq:idEhoney}
\end{equation}
we achieve a flat band for $\beta=0$ or $\kappa=0$.
By comparing two characteristic polynomials for Eqs.(\ref{eq:1Hhoney}) and (\ref{eq:idEhoney}), 
we obtain five equations,
\begin{align}
&a+b=\alpha + \gamma,
\nonumber\\
&d+e= \beta+\kappa,
\nonumber\\
&ab-3c^{2}= \alpha\gamma,
\nonumber\\
&ae+bd-c^2 = \alpha\kappa + \beta \gamma,
\nonumber\\
&de=\beta\kappa\,,
\end{align}
which lead to a condition for $a,b,c,d,e$ for the honeycomb model as
\begin{equation}
c^{2}\,=\, (d-e)[3(d-e)-(a-b)]\,.
\label{eq:fbc_honey}
\end{equation}
Again, this auxiliary condition is a sufficient 
condition for the model to have a parameter region for flat bands, 
which we impose hereafter, so that 
the model contains four independent parameters.

By diagonalizing the Hamiltonian matrix (\ref{eq:1Hhoney}) with the auxiliary condition (\ref{eq:fbc_honey}),
we obtain the upper and lower energy bands,
\begin{align}
E^{\pm}=
\begin{cases}
b+3d-3e \,+\, d\,F^{\rm honeycomb}(k_{x},k_{y}) \\
a+3e-3d \,+\,e\,F^{\rm honeycomb}(k_{x},k_{y}),
\end{cases}
\label{eq:Ehoney}
\end{align}
where the parameter $c$ is eliminated with use of the condition.
It is obvious that $d=0$ or $e=0$ yields a flat band.
Again, the energy of the flat band is nonzero in general, 
so that this model is not necessarily the rank-reduction type. 
Equation~(\ref{eq:Ehoney}) indicates that the gap between the flat and dispersive bands is 
given by $b-a+3d$ for $e=0$ or $a-b+3e$ for $d=0$, which are tunable.
As in the tetragonal case, 
setting $|c|=1$ and $d=0$ or $e=0$ leaves us two adjustable parameters, which enable us to control both the energy of the flat band and the gap from the dispersive band.

\begin{figure}[htbp]
\begin{center}
\includegraphics[width=0.48\textwidth]{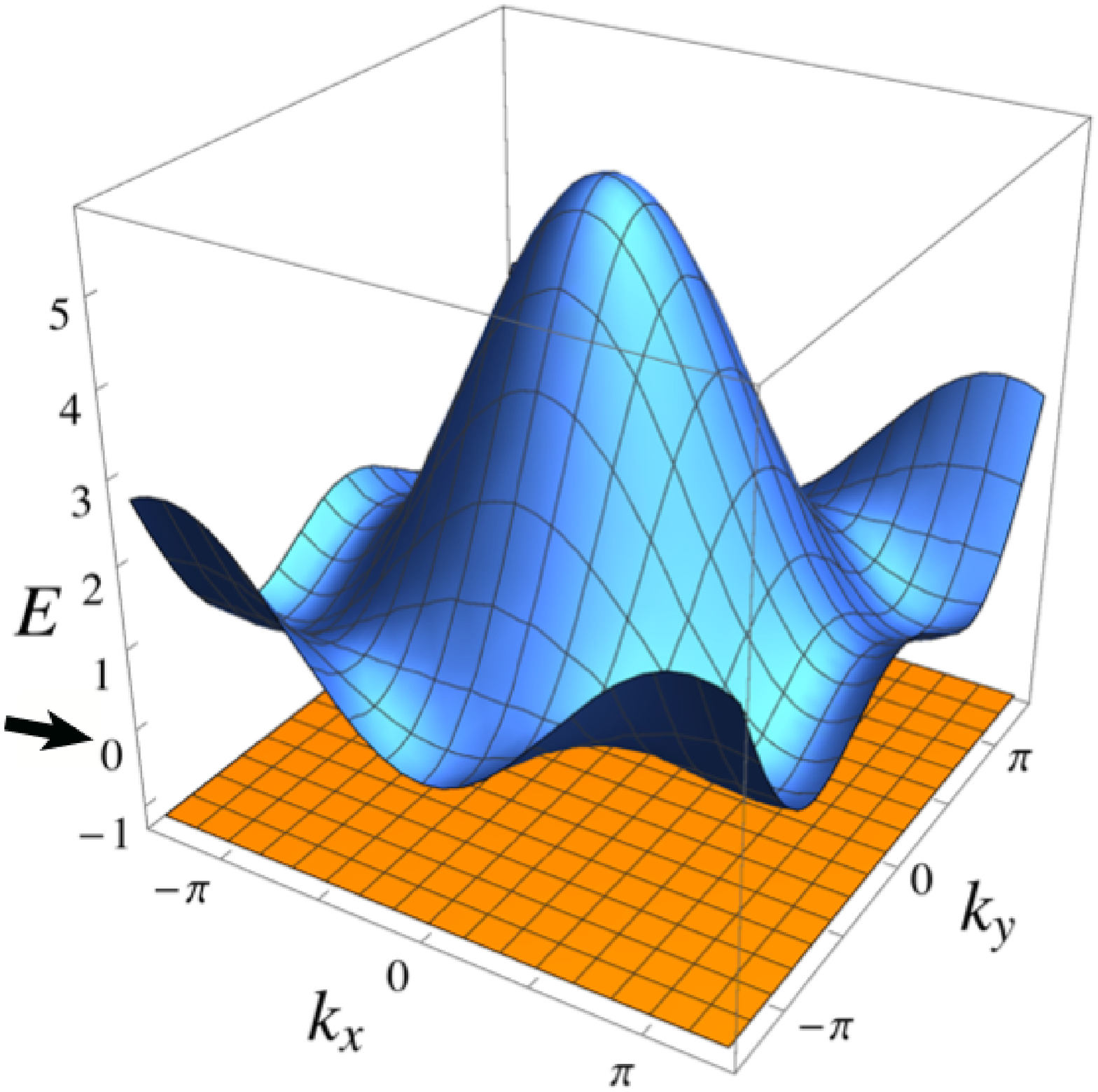}
\end{center}
\caption{The band structure of the non-rank-reduced type 
honeycomb flat-band model for $(a,b,c,d,e)=(1/\sqrt{3},1/\sqrt{3},1,1/\sqrt{3},0)$. 
For honeycomb lattices, energy bands are depicted in an extended zone 
scheme, while the hexagonal first Brillouin zone is surrounded by $K$ and $K'$ points at $(k_{x},k_{y})=({2\pi\over{\sqrt{3}}},\pm{2\pi\over{3}})(\pm{2\pi\over{\sqrt{3}}},{2\pi\over{3}})(0,\pm{4\pi\over{3}})$.
}
\label{fig:honey01}
\end{figure}

{\it Non-rank-reduced honeycomb model}: 
In Fig.~\ref{fig:honey01}, we depict a typical example of the non-rank-reduced honeycomb flat-band model
for $(a,b,c,d,e)=(1/\sqrt{3},1/\sqrt{3},1,1/\sqrt{3},0)$, where the flat band has a nonzero energy.  

\begin{figure}[htbp]
\begin{center}
\includegraphics[width=0.48\textwidth]{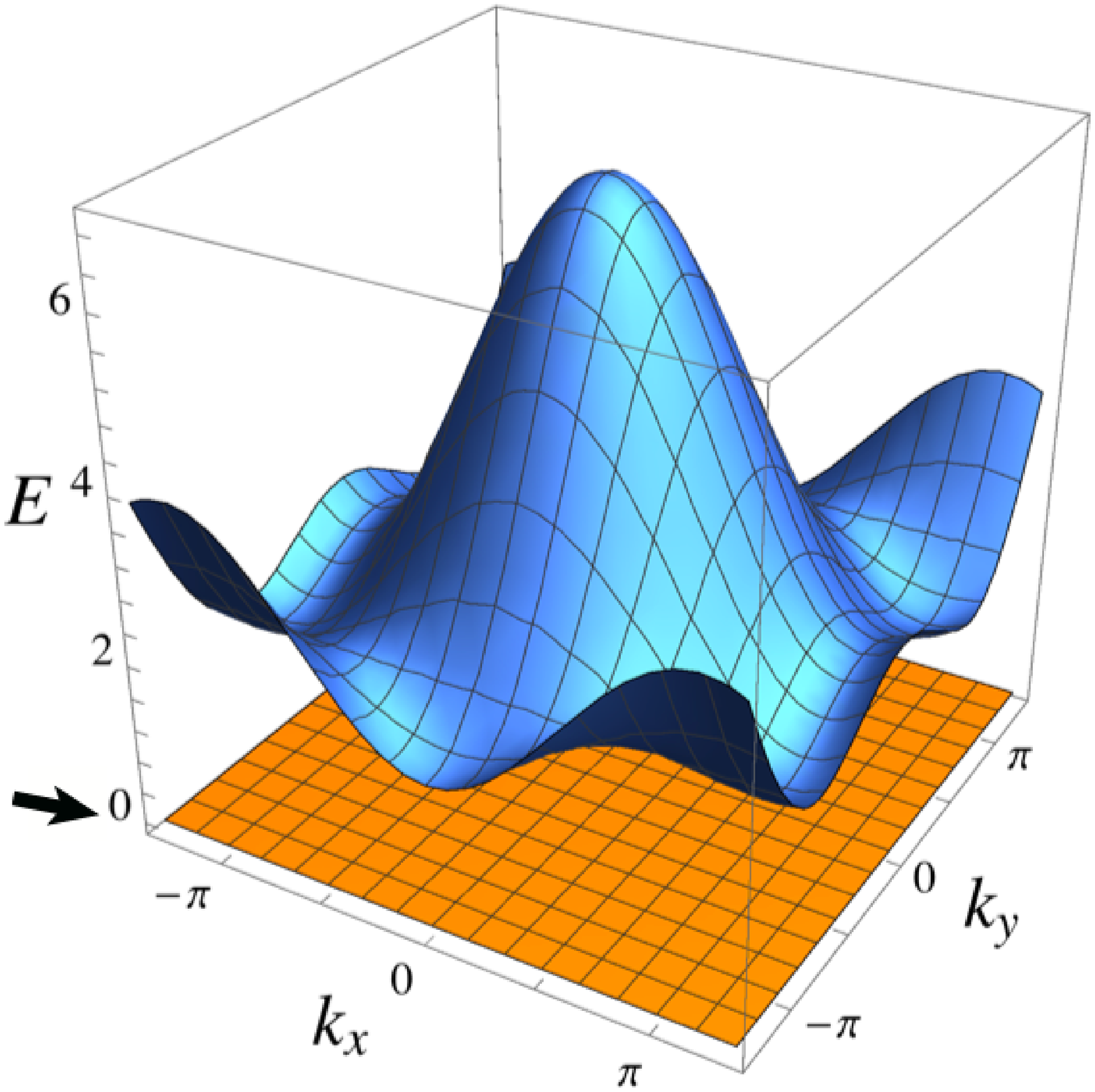}
\end{center}
\caption{The energy band structure of the rank-reduced type honeycomb flat-band model for $(a,b,c,d,e)=(\sqrt{3},\sqrt{3},1,1/\sqrt{3},0)$ .
}
\label{fig:honey02}
\end{figure}

{\it Rank-reduced  honeycomb model}: 
The rank-reduced type of the honeycomb flat-band model is obtained by imposing
a further condition,
\begin{equation}
b+3d-3e =0\quad\quad {\rm or}\quad\quad a+3e-3d=0 \,,
\label{eq:rrc_honey}
\end{equation}
in which the wavenumber-independent part of upper or lower bands becomes zero.
A typical  case is depicted in Fig.~\ref{fig:honey02} for 
$(a,b,c,d,e)=(\sqrt{3},\sqrt{3},1,1/\sqrt{3},0)$, 
which satisfies the rank-reduction conditions (\ref{eq:rrc_honey}), where 
the flat band indeed has exactly zero energy.

Another type of the rank-reduced honeycomb models is obtained
by imposing that the wavenumber-independent part of the eigenvalues becomes zero simultaneously with the coefficients of the wavenumber-dependent part ($d$ or $e$), i.e.,
\begin{align}
&b+3d-3e\to0 \,\,\;{\rm for}\;\,\, d=0
\nonumber\\
&\quad\quad\quad\quad\quad {\rm or}
\nonumber\\
&a+3e-3d\to0\,\, \;{\rm for}\; \,\,e=0\,.
 \label{eq:honey_tasaki}
\end{align}
If this condition is satisfied, the relevant eigenvalues become zero. 
The models with this condition contain Tasaki's $(n,m)=(3,3)$ model \cite{Tasaki:1992}, 
where the parameters $\nu, t, s$ defining the 
model is related to the present one as
\begin{align}
&a=3t\nu^{2}-s ,\quad b=t-3s\nu^{2} \,,
\nonumber\\
&c=\nu(t+s),\quad d=\nu^{2}t, \quad e=-\nu^{2} s\,.
\label{eq:tasaki_honey}
\end{align}
This parametrization leads to
\begin{align}
&b+3d-3e \,=\,(1+3\nu^2 )t, \\
&a+3e-3d\,=\,-(1+3\nu^2 )s\,.
\end{align}
Since $d,e$ and $t,s$ are related as $d=\nu^2 t$ and $e=-\nu^2 s$, 
the condition (\ref{eq:honey_tasaki}) is satisfied for $\nu\not=0$.
Here, setting $d=0$ or $e=0$ (equivalently $t=0$ or $s=0$) results in exactly zero-energy flat band. 
We again have one adjustable parameter for a fixed unit of energy ($|c|=|\nu(t+s)|=1$). Similar to the tetragonal case, 
we can infer that the connectivity condition is satisfied with the Wannier orbitals unorthogonalizable 
in the present honeycomb model, as in Tasaki's $(n,m)=(3,3)$ model.

We note that a related honeycomb flat-band model with no onsite potentials is 
discussed in Ref.\cite{Raoux:2017rao}.


\section{Flat-band model on 3D bcc lattices}
\label{sec:bcc}

\begin{figure}[htbp]
\begin{center}
\includegraphics[width=0.5\textwidth]{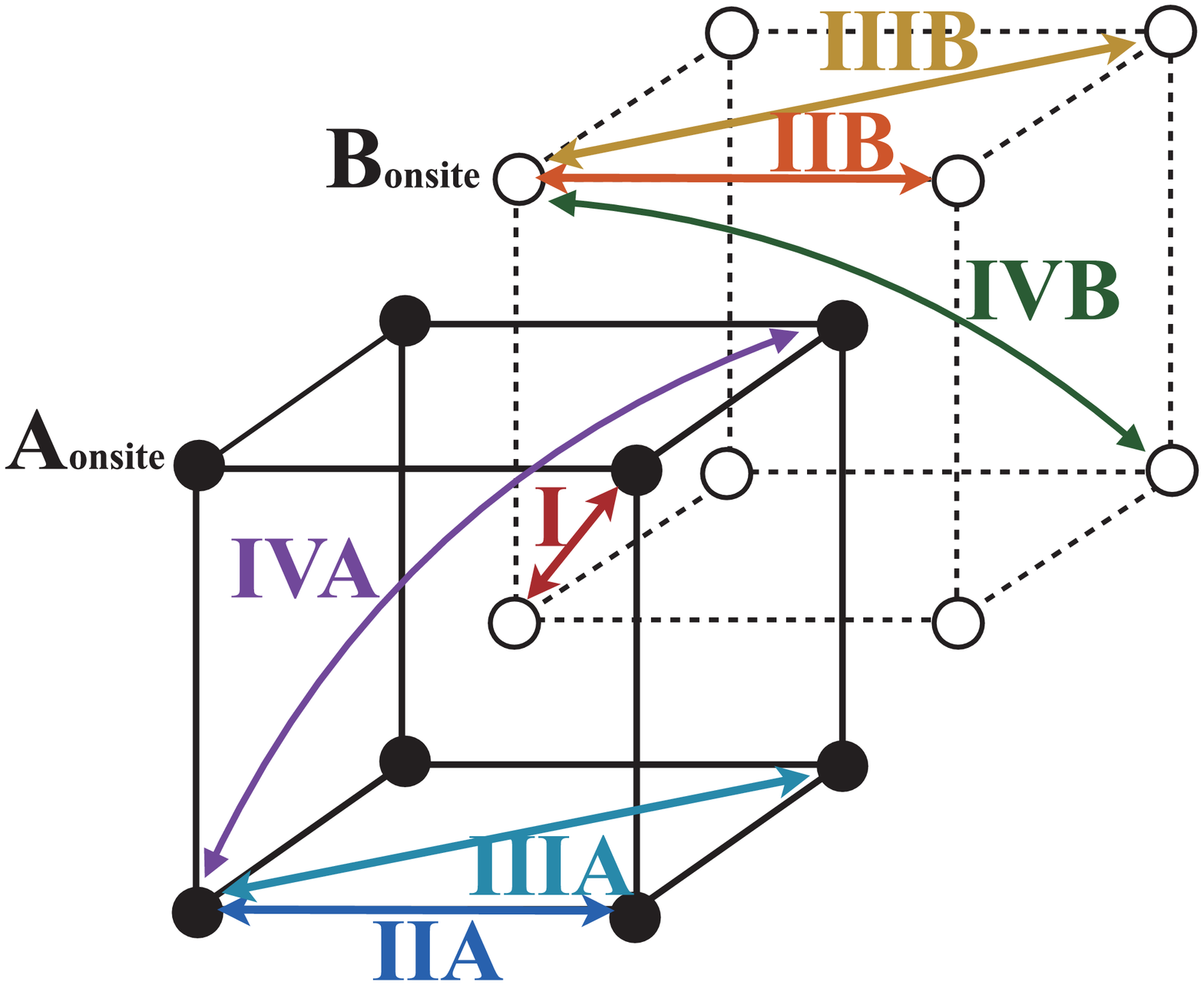}
\end{center}
\caption{The 3D bcc lattice, where 
the hoppings (arrows) are given by Eq.(\ref{eq:bcc}) in the text. 
Solid and open circles represent A and B sublattice sites, respectively.}
\label{fig:3Dsites}
\end{figure}

The formulation for 2D lattices above 
is so general that we can readily extend the tetragonal model to three-dimensional 
body-centered cubic (bcc) lattices, as depicted in Fig.~\ref{fig:3Dsites}.  
There, we need to consider the nearest ($t$), second-neighbor 
($t^{(2)}$), third-neighbor ($t^{(3)}$) and fourth-neighbor ($t^{(4)}$) 
hoppings along with onsite energies with five parameters ($a,b,c,d,e$) 
as
\begin{align}
&{\rm A_{onsite}}\,\,:\,\, t_{\rm AA}^{(0)} = a,\quad
{\rm B_{onsite}}\,\,:\,\, t_{\rm BB}^{(0)} = b, \nonumber\\
&{\rm I}\,\,:\,\,\quad\,t_{\rm AB} = c, \nonumber\\
&{\rm IIA}\,:\,t_{\rm AA}^{(2)} = d,\,\,\,\,
{\rm IIIA}\,:\, t_{\rm AA}^{(3)} = {d\over2},\,\,\,\,
{\rm IVA}\,:\, t_{\rm AA}^{(4)} = {d\over4}, \nonumber\\
&{\rm IIB}\,:\, t_{\rm BB}^{(2)} = e,\,\,\,\,
{\rm IIIB}\, : \, t_{\rm BB}^{(3)} = {e\over2},\,\,\,\, 
{\rm IVB}\, : \, t_{\rm BB}^{(4)} = {e\over4}, \nonumber\\
\label{eq:bcc}
\end{align}
where subscripts denote sublattices and $a,b,c,d,e \in {\mathbb R}$.  
For the hoppings on each of A and B sublattices, 
the next-nearest (diagonal on a plane) 
hopping on A(B) sublattice is set to be half 
the nearest one, $d(e)$, while the next-next-nearest one 
(diagonal on a cube) is set 
to be half the next-nearest one.  

In $k$-space the Hamiltonian is
\begin{align}
{\mathcal H}=
\begin{pmatrix}
{\mathcal H}_{\rm AA}  & {\mathcal H}_{\rm AB} \\
{\mathcal H}_{\rm BA} & {\mathcal H}_{\rm BB}
\end{pmatrix}\,,
\label{eq:1Hbcc}
\end{align}
with
\begin{align}
&{\mathcal H}_{\rm AA}  = a+  2dF(k_{x},k_{y},k_{z})\,,
\nonumber\\
&{\mathcal H}_{\rm AB} = {\mathcal H}_{\rm BA} = 8c\cos{k_{x}\over{2}}\cos {k_{y}\over{2}}\cos {k_{z}\over{2}} \,,
\nonumber\\
&{\mathcal H}_{\rm BB} = b+  2eF(k_{x},k_{y},k_{z})\,,
\label{eq:1Hbcc2}
\end{align}
where we have defined
\begin{align}
F(k_{x},k_{y},k_{z})&\equiv \cos k_{x}+\cos k_{y} + \cos k_{z} 
\nonumber\\
&+ \cos k_{x} \cos k_{y} +\cos k_{y} \cos k_{z}+ \cos k_{z} \cos k_{x}
\nonumber\\
&+\cos k_{x} \cos k_{y} \cos k_{z}\,.
\end{align}

Desirable forms of energy eigenvalues are 
\begin{equation}
E=
\begin{cases}
\alpha + \beta F(k_{x},k_{y},k_{z})\\
\gamma + \kappa F(k_{x},k_{y},k_{z})
\end{cases},
\label{eq:idE_bcc}
\end{equation}
where we achieve a flat band when $\beta=0$ or $\kappa=0$.  
We again consider the characteristic polynomials for Eqs.(\ref{eq:1Hbcc}),(\ref{eq:idE_bcc}), where the former becomes
\begin{align}
&E^2 -\Big[(a+b)+2(d+e)F(k_{x},k_{y},k_{z})\Big]E
\nonumber\\
&+ab+2(ae+bd)F(k_{x},k_{y},k_{z})
+4de\,F(k_{x},k_{y},k_{z})^2
\nonumber\\
&-8c^{2}[1+F(k_{x},k_{y},k_{z})]=0\,,
\label{3D1}
\end{align}
while the latter is
\begin{align}
&E^2 -\Big[(\alpha+\gamma)+(\beta+\kappa)F(k_{x},k_{y},k_{z})\Big]E
\nonumber\\
&+\alpha\gamma+(\alpha\kappa+\beta\gamma)F(k_{x},k_{y},k_{z})
+\beta\kappa \,F(k_{x},k_{y},k_{z})^2=0\,.
{\label{3D2}}
\end{align}
By comparing the two characteristic polynomials,
we obtain five equations,
\begin{align}
&a+b=\alpha + \gamma,
\nonumber\\
&2(d+e)= \beta+\kappa,
\nonumber\\
&ab-8c^{2}= \alpha\gamma,
\nonumber\\
&2ae+2bd-8c^2 = \alpha\kappa + \beta \gamma,
\nonumber\\
&4de=\beta\kappa.
\end{align}
From these we eliminate $\alpha,\beta,\gamma,\kappa$ 
to obtain a single equation for $a,b,c,d,e$, which amounts to 
an auxiliary condition,
\begin{equation}
c^{2}\,=\, {2(d-e)^{2} -(d-e)(a-b) \over{4}}\,.
\label{eq:fbc_bcc}
\end{equation}
Imposing this, the model 
is left with four independent parameters.

By diagonalizing the Hamiltonian (\ref{eq:1Hbcc}) with the auxiliary condition (\ref{eq:fbc_bcc}), 
we obtain the upper and lower energy bands as
\begin{align}
E^{\pm}=
\begin{cases}
b+2d-2e + 2d\,F(k_{x},k_{y},k_{z}) \\
a+2e-2d + 2e\,F(k_{x},k_{y},k_{z})\,,
\end{cases}
\label{eq:Ebcc}
\end{align}
where the parameter $c$ is eliminated through the auxiliary condition.
Obviously $d=0$ or $e=0$ yields a flat band.  
The models are extended to higher dimensions in Appendix.~\ref{sec:bccD}.


\section{Summary and Discussions}
\label{sec:SD}

In this work, we have proposed flat-band models on the 2D fcc lattice and the honeycomb lattice.
We have considered the tight-binding models with up to the third-neighbor hoppings, for which we impose an auxiliary condition for a dispersionless band to emerge.  
We have then succeeded in obtaining flat-band models in two classes: 
one class has a flat band separated in energy from a dispersive band, 
where the present model has virtues of a nonzero flat-band energy and a tunable gap between the bands.  
The other class, obtained by deforming the first class, 
has a (nearly) flat band that pierces right through a dispersive band.  
So this is a first realization of a two-dimensional flat band 
crossing with a parabolic band in two-band models.  
In both classes we have in general non-rank-reduced situation that 
contains rank-reduced one as a special case.  
We have noted that the present models include Tasaki's cell-constructed flat-band models.  
We have also indicated that these models satisfy the connectivity condition 
as in the known flat-band models, with
the Wannier orbitals unorthogonalizable.  
We have shown that we can construct higher-dimensional versions of the flat-band models by introducing more distant hoppings.

Future works should include the following: 
(i) Study of superconductivity in the deformed flat-band model with the intersection of the flat and dispersive bands to ascertain whether 
the pair scatterings between the dispersive and flat bands can really 
enhance superconductivity.  
(ii) To elucidate topological properties of the flat-band systems. In our 
(undeformed) model a flat band has a nonzero energy and gapped from the dispersive band, for which we can define a Chern number to discuss topological properties of the flat band. Refs.\cite{Tang:2011tmw,Sun:2011sgk,Neupert:2011nsc,Li:2013lzl} indeed report that the flat-band systems can realize
topological insulators with nontrivial Chern numbers. Superfluid weight and its relation with the topological number is another important property as investigated in Ref.\cite{Tovmasyan:2016:tpt}.
On the other hand, Ref.\cite{Chen:2014cms} argues that 
three conditions --- finite-ranged hoppings, nonzero Chern number, 
and exactly flat band --- cannot be simultaneously satisfied.  
The present deformed models in the present scheme have 
warped flat bands, so that it is an interesting question 
as to whether this allows nonzero Chern numbers.  
(iii) We also would like to see whether the present scheme for construction of the bipartite, two-band flat-band models may be extended to three-band 
and higher number of bands.


\begin{acknowledgments}
The authors wish to thank K.~Kuroki and D.~Ogura for illuminating 
discussions, especially on superconductivity in flat-band systems.  
The authors also thank T.~Oka and T.~Morimoto for discussions 
in an early stage of this work.  
H.A. was supported by JSPS KAKENHI Grant No.
26247057 and ImPACT Program of Council for Science, Technology and Innovation, Cabinet Office, Government of Japan (Grant No. 2015-PM12-05-01) from JST. 
T.M. was supported by JSPS KAKENHI Grant No.16K17677, 
and also by MEXT-Supported Program for the Strategic Research Foundation
at Private Universities (Keio University) ``Topological Science" (Grant No. S1511006).

\end{acknowledgments}


\appendix

\section{Tasaki's $(n,m)=(4,4)$ Model}
\label{sec:DTM}

Hopping parameters in the Tasaki $(n,m)=(4,4)$ model are given by
\begin{align}
&{\rm I}\,\,:\,\,t_{\rm AB}=\nu(t+s), \nonumber\\
&{\rm II}\,\,:\,\,t_{\rm AA}^{(1)}=2\nu^2 t, \quad 
&{\rm III}\,\,:\,\, t_{\rm AA}^{(2)}=\nu^2 t, \nonumber\\
&{\rm IV}\,\,:\,\, t_{\rm BB}^{(1)}=-2\nu^2 s ,\quad
&{\rm V}\,\, : \,\, t_{\rm BB}^{(2)}=-\nu^2 s, \nonumber\\
&{\rm A}\,\,:\,\, t_{\rm AA}^{(0)}=4t\nu^2 -s ,\quad
&{\rm B}\,\,:\,\, t_{\rm BB}^{(0)}=t-4s\nu^2  \,.
\end{align}
The Hamiltonian in $k$-space is given by
\begin{align}
&{\mathcal H}_{\rm AA}  = - s+ 4t\nu^2 (1+\cos k_{x}+\cos k_{y} + \cos k_{x} \cos k_{y}) \,,
\nonumber\\
&{\mathcal H}_{\rm AB} ={\mathcal H}_{\rm BA}= 4\nu(t+s)\cos{k_{x}\over{2}}\cos {k_{y}\over{2}} \,,
\nonumber\\
&{\mathcal H}_{\rm BB} = t- 4s\nu^2 (1+\cos k_{x}+\cos k_{y} + \cos k_{x}) \,.
\end{align}

\section{Tasaki's $(n,m)=(3,3)$ Model}
\label{sec:t33}

Hopping parameters in the Tasaki $(n,m)=(3,3)$ model
are given by
\begin{align}
&{\rm I}\,\,:\,\,t_{\rm AB} = \nu(t+s), \nonumber\\
&{\rm II}\,\,:\,\,t_{\rm AA}^{(1)} = \nu^{2}t,\quad
&{\rm III}\,\,:\,\, t_{\rm BB}^{(1)} = -\nu^{2} s,\nonumber\\
&{\rm A}\,\,:\,\, t_{\rm AA}^{(0)} = 3t\nu^{2}-s,\quad
&{\rm B}\,\,:\,\, t_{\rm BB}^{(0)} = t-3s\nu^{2} \,.
\end{align}
The Hamiltonian in $k$-space  is given by
\begin{align}
&{\mathcal H}_{\rm AA}  = 3t\nu^{2}-s+  \nu^{2}t\left(4\cos {\sqrt{3}k_{x}\over{2}}\cos {k_{y}\over{2}} +  2\cos k_{y}\right)\,,
\nonumber\\
&{\mathcal H}_{\rm AB} = {\mathcal H}_{\rm BA}^{*} 
\nonumber\\
&= \nu(t+s)\left[e^{i{k_{x}\over{\sqrt{3}}}} 
+ e^{i\left( - {\sqrt{3}k_{x}\over{6}}+{k_{y}\over{2}}\right)}+ e^{i\left( - {\sqrt{3}k_{x}\over{6}}-{k_{y}\over{2}}\right)}\right],
\nonumber\\
&{\mathcal H}_{\rm BB} = t-3s\nu^{2}-\nu^{2} s\left(4\cos {\sqrt{3}k_{x}\over{2}}\cos {k_{y}\over{2}} +  2\cos k_{y}\right)\,.
\end{align}

\section{Flat-band model on general-dimensional bcc lattices}
\label{sec:bccD}

We can extend the body-centered tetragonal models discussed in the main text to general $D$-dimensional bcc lattices, where we consider the hoppings,
\begin{align}
&t_{\rm AA}^{(0)} = a,\quad
t_{\rm BB}^{(0)} = b, \nonumber\\
&t_{\rm AB} = c, \nonumber\\
&t_{\rm AA}^{(2)} = d,\quad
t_{\rm AA}^{(3)} = {d\over2},\,\,\cdot\cdot\cdot,\,\,
t_{\rm AA}^{(D+1)} = {d\over{2^{D-1}}}, \nonumber\\
&t_{\rm BB}^{(2)} = e,\quad
t_{\rm BB}^{(3)} = {e\over2},\,\, \cdot\cdot\cdot,\,\,
 t_{\rm BB}^{(D+1)} = {e\over{2^{D-1}}}, \nonumber\\
\label{eq:bccD}
\end{align}
with subscripts denoting sublattices and $a,b,c,d,e \in {\mathbb R}$.
Here, $c$ characterizes the nearest hopping between AB, 
while $d(e)$ characterize AA (BB) hoppings, where 
the $(n+1)$-th neighbor hopping on each (A or B) sublattice is set to be 
half the $n$-th neighbor one on the sublattice.  
In passing, we can note that this is something like 
a finite-ranged version of a model 
studied in Ref.\cite{Neupert:2011nsc} where the (infinite-ranged) 
hopping amplitude decays exponentially with distance.

As in 2D and 3D cases, we compare the characteristic polynomials for (\ref{eq:bccD}) and the desirable form of energy eigenvalues obtained by replacing $F(k_{x}, k_{y}, k_{z})$ in Eqs.(\ref{3D1}),(\ref{3D2}) by
\begin{align}
&F(k_{1},k_{2},...,k_{D})=\prod_{i=1}^{D}(1+\cos k_{i})\,\,-\,\,1,
\end{align} 
and replacing a factor $8$ of the last term on the left-hand side of Eq.(\ref{3D1}) by $2^{D}$.
We can then obtain five equations similar to those in the 3D case as
\begin{align}
&a+b=\alpha + \gamma,
\nonumber\\
&2(d+e)= \beta+\kappa,
\nonumber\\
&ab-2^{D}c^{2}= \alpha\gamma,
\nonumber\\
&2ae+2bd-2^{D}c^2 = \alpha\kappa + \beta \gamma,
\nonumber\\
&4de=\beta\kappa.
\end{align}
From these we eliminate $\alpha,\beta,\gamma,\kappa$ 
to obtain am equation for $a,b,c,d,e$, 
\begin{equation}
c^{2}\,=\, {2(d-e)^{2} -(d-e)(a-b) \over{2^{D-1}}}\,.
\label{eq:fbc_bccD}
\end{equation}

The energy eigenvalues for the upper and lower energy bands are
\begin{align}
E^{\pm}=
\begin{cases}
b+2d-2e + 2d\,F(k_{1},k_{2},...,k_{D})  \\
a+2e-2d + 2e\,F(k_{1},k_{2},...,k_{D}) \,,
\end{cases}
\label{eq:EbccD}
\end{align}
where the parameter $c$ is eliminated through the auxiliary condition.
Again, $d=0$ or $e=0$ yields a flat band.


\end{document}